\documentclass[12pt,reqno]{amsart}

\usepackage{amssymb,amsthm,amsfonts,latexsym}
\usepackage{amsmath}
\usepackage{mathtools}
\usepackage{enumitem}
\usepackage{needspace}
\usepackage{array}
\usepackage{graphicx,booktabs,tabularx,longtable,pdflscape,float}
\usepackage{xurl}
\usepackage[multiple]{footmisc}
\usepackage{comment}

\usepackage[authoryear,round]{natbib}

\usepackage[letterpaper,left=1.25in,top=1.25in,right=1.25in,bottom=1.25in]{geometry}
\usepackage{setspace}
\usepackage[T1]{fontenc}
\usepackage{tgpagella}
\usepackage{microtype}

\usepackage[dvipsnames]{xcolor}
\usepackage{tikz}
\usepackage{ifthen}

\usepackage{etoolbox}
\pretocmd{\subsubsection}{\needspace{5\baselineskip}}{}{}
\pretocmd{\subsection}{\needspace{3\baselineskip}}{}{}
\patchcmd{\section}{\scshape}{\bfseries}{}{%
  \PackageWarning{exchange}{Section-heading font patch failed}}
\makeatletter
\renewcommand{\@secnumfont}{\bfseries}
\makeatother

\theoremstyle{plain}
\newtheorem{theorem}{Theorem}
\newtheorem{corollary}{Corollary}
\newtheorem{lemma}{Lemma}
\newtheorem{proposition}{Proposition}

\theoremstyle{definition}
\newtheorem{definition}{Definition}

\newtheoremstyle{boldremark}
 {0.5\topsep}{0.5\topsep}
 {\normalfont}{}
 {\bfseries}{.}
 {.5em}{}
\theoremstyle{boldremark}
\newtheorem{remark}{Remark}

\usepackage[
 colorlinks=true,
 linkcolor=blue,
 citecolor=blue,
 urlcolor=blue,
 filecolor=blue,
 hyperfootnotes=false,
 hyperindex,
 breaklinks
]{hyperref}

\renewcommand{\sectionautorefname}{Section}
\renewcommand{\subsectionautorefname}{Section}
\renewcommand{\subsubsectionautorefname}{Section}

\newcommand{\Acore}{\mathcal C}
\setlist{itemsep=3pt,topsep=5pt}

\newif\ifsupplement
\supplementtrue 
\newcommand{\declaresuppref}[3]{%
  \csdef{suppname:#1}{#2}\csdef{suppnumber:#1}{#3}%
  \ifsupplement
    \AtEndDocument{\ifcsdef{r@#1}{%
      \edef\currentsuppnumber{\getrefnumber{#1}}%
      \ifdefstring{\currentsuppnumber}{#3}{}{%
        \PackageWarningNoLine{exchange}{Update standalone number for #1 to \currentsuppnumber}}%
    }{}}%
  \fi}
\declaresuppref{app:robust-proofs}{Appendix}{S.1}
\declaresuppref{app:compatibility-uniqueness}{Appendix}{S.2}
\declaresuppref{app:priority-free}{Appendix}{S.3}
\declaresuppref{app:policy}{Appendix}{S.4}
\declaresuppref{app:chains}{Appendix}{S.5}
\declaresuppref{fig:policy-cycle}{Figure}{S.1}
\declaresuppref{prop:priority-free}{Proposition}{S.1}
\declaresuppref{prop:certifying}{Proposition}{S.2}
\declaresuppref{prop:reuse}{Proposition}{S.3}
\declaresuppref{prop:released-priority}{Proposition}{S.3}

\NewCommandCopy{\standardautoref}{\autoref}
\RenewDocumentCommand{\autoref}{s m}{%
  \ifcsdef{suppname:#2}{%
    \ifsupplement
      \IfBooleanTF{#1}{\csuse{suppname:#2}~\ref*{#2}}%
        {\hyperref[#2]{\csuse{suppname:#2}~\ref*{#2}}}%
    \else
      \csuse{suppname:#2}~\csuse{suppnumber:#2}%
    \fi
  }{\IfBooleanTF{#1}{\standardautoref*{#2}}{\standardautoref{#2}}}}
\newcommand{\suppref}[2][]{\autoref{#2}#1}
\newcommand{\bodysecref}[1]{\begingroup
  \renewcommand{\sectionautorefname}{Section}%
  \renewcommand{\subsectionautorefname}{Section}%
  \renewcommand{\subsubsectionautorefname}{Section}%
  \autoref{#1}\endgroup}
\newcommand{\appref}[1]{\begingroup
  \renewcommand{\sectionautorefname}{Appendix}%
  \renewcommand{\subsectionautorefname}{Appendix}%
  \renewcommand{\subsubsectionautorefname}{Appendix}%
  \renewcommand{\appendixautorefname}{Appendix}%
  \autoref{#1}\endgroup}

\hypersetup{pdftitle={Exchange rights and vacancy priorities},
 pdfauthor={Özgün Ekici and M. Bumin Yenmez}}
\title{Exchange rights and vacancy priorities}
\thanks{\"Ozg\"un Ekici: Department of Economics, \"Ozye\u{g}in University, Istanbul, T\"urkiye. Email: \texttt{ozgun.ekici@ozyegin.edu.tr}. M. Bumin Yenmez: Department of Economics, Washington University in St. Louis, USA; Durham University, UK; and \"Ozye\u{g}in University, T\"urkiye. Email: \texttt{bumin@wustl.edu}. AI disclosure: The authors used ChatGPT and Claude to assist with manuscript preparation, including exposition, LaTeX editing, and checking mathematical arguments. The authors take full responsibility for the content of the paper, including its results, proofs, and references.}
\author[Ö. Ekici]{\"Ozg\"un Ekici}
\author[M. B. Yenmez]{M. Bumin Yenmez}
\date{}

\begin{document}
\begin{abstract}
Many institutions let incumbents exchange positions while allocating
announced vacancies by priority, withdrawing positions released by
vacancy recipients. A policy specifies permitted exchanges,
vacancy eligibility, and priorities. A strategy-proof, individually
rational rule protecting exchange rights and vacancy priorities exists
precisely for policies without conflict cycles. Restricting protection
to first choices does not enlarge this class. We introduce the
propose-and-trade rule, which satisfies these properties and selects
the applicant-optimal core allocation under every conflict-free policy.
We identify when propose-and-trade is the unique strategy-proof,
individually rational rule protecting first choices. These axioms
characterize deferred acceptance and top trading cycles in their
classic settings. Under a policy with a conflict cycle, propose-and-trade
is manipulable but protects exchange rights and first-choice claims.
Replacing it with any strategy-proof rule makes some applicant worse
off at some profile. We show when priority reform alone can eliminate
conflict cycles and when eligibility or exchange permissions must
change.


\end{abstract}
\maketitle

\section{Introduction}\label{sec:intro}
\enlargethispage{-6pt} 

Public employers and housing authorities often operate two allocation channels side by side. In one, incumbents and newcomers compete for announced vacancies by priority.\footnote{Web pages cited in this paper were accessed in November 2026. Newcomers participate, for example, as probationary teachers seeking their first permanent placement in Spain: see \href{https://www.boe.es/eli/es/rd/2010/10/29/1364/con}{Royal Decree 1364/2010 of 29 October}, regulating nationwide teacher-transfer competitions, consolidated text of 17 July 2024, Art.~13. This competition also includes released positions when educational planning provides for their continuation (\autoref{sec:chains}).} In the other, incumbents may also exchange the positions they hold. Each channel alone has a well-understood strategy-proof rule: deferred acceptance
(DA) respects vacancy priorities \citep{GS1962}, and top trading cycles
(TTC) respects exchange rights \citep{SS1974}. We ask which institutional \emph{policies}, specifying permitted exchanges, eligibility, and priorities, allow a single strategy-proof rule to integrate the two channels while respecting both exchange rights and vacancy priorities. When no such rule exists, we ask how the policy must be reformed to allow such a rule.

In practice, both channels are often governed by laws and regulations. In English social housing, for instance, tenants whose transfer requests qualify for statutory preference are considered alongside new applicants, while tenants with secure tenancies may also exchange homes with other eligible social tenants. Landlords may refuse such exchanges only on statutory grounds. In Spain's central government, eligible permanent labor staff compete by priority for published posts and may also request reciprocal transfers. In Turkey's Ministry of Family and Social Services, eligible employees compete by priority for advertised posts and may also request reciprocal transfers, or \emph{becayi\c{s}}. In Colombia's public schools, ordinary transfer rounds allocate published vacancies using announced selection criteria, and teachers may also request mutual exchanges, or \emph{permutas}. In Algeria, teachers apply for vacant posts in a transfer round and may subsequently exchange posts. In each case, incumbents can both compete for vacancies and exchange their positions.\footnote{For England, see the \href{https://www.legislation.gov.uk/ukpga/1996/52/section/159}{Housing Act 1996}, ss.~159 and~166A, and the \href{https://www.legislation.gov.uk/ukpga/1985/68/section/92}{Housing Act 1985}, s.~92 and Sch.~3. Cheltenham, for example, offers both channels: see Cheltenham Borough Council, \href{https://www.cheltenham.gov.uk/housing/cheltenham-borough-council-housing-services/your-tenancy/swapping-and-transferring-your-home/}{\emph{Swapping and transferring your home}}. For Spain, see the Directorate-General for the Civil Service, \href{https://www.boe.es/buscar/doc.php?id=BOE-A-2023-26762}{Resolution of 28 December 2023 regulating the open and permanent transfer competition}, which covers permanent labor staff under the IV Convenio \'Unico. For Turkey, see the ministry's \href{https://resmigazete.gov.tr/eskiler/2013/05/20130517-12.htm}{2013 Regulation on appointments and transfers}, Arts.~11, 20, and~22, and its annual internal-transfer notices, such as the \href{https://www.aile.gov.tr/media/301285/2026-yer-degisikligi-duyurusu.pdf}{2026 notice}. For Colombia, see \href{https://www.sucop.gov.co/entidades/mineducacion/Paginas/decreto1075de2015.aspx}{Decree 1075 of 2015}, Arts.~2.4.5.1.2 to~2.4.5.1.4 for ordinary transfers and Art.~2.4.5.1.2, par\'agrafo~2, for mutual exchanges. For Algeria, see the Ministry of National Education's \href{https://www.education.gov.dz/\%D8\%A7\%D9\%84\%D8\%B1\%D8\%B2\%D9\%86\%D8\%A7\%D9\%85\%D8\%A9/}{2026 transfer calendar} and an Alg\'erie Presse Service \href{https://www.aps.dz/algerie/education-et-technologie/mtu61np6-\%D8\%A7\%D9\%86\%D8\%B7\%D9\%84\%D8\%A7\%D9\%82-\%D8\%B9\%D9\%85\%D9\%84\%D9\%8A\%D8\%A9-\%D9\%86\%D9\%82\%D9\%84-\%D8\%A7\%D9\%84\%D8\%A7\%D9\%94\%D8\%B3\%D8\%A7\%D8\%AA\%D8\%B0\%D8\%A9-\%D8\%AF\%D8\%A7\%D8\%AE\%D9\%84-\%D9\%88\%D8\%AE\%D8\%A7\%D8\%B1\%D8\%AC-\%D8\%A7\%D9\%84\%D9\%88\%D9\%84\%D8\%A7\%D9\%8A\%D8\%A9-\%D8\%B9\%D9\%86-\%D8\%B7\%D8\%B1\%D9\%8A\%D9\%82-\%D8\%A7\%D9\%84\%D8\%AA\%D8\%A8\%D8\%A7\%D8\%AF\%D9\%84-\%D8\%A7\%D8\%A8\%D8\%AA\%D8\%AF\%D8\%A7\%D8\%A1-\%D9\%85\%D9\%86-\%D8\%A7\%D9\%84\%D9\%8A\%D9\%88\%D9\%85}{report of 9 September 2026 on the opening of exchanges} (both in Arabic).}

In our model an incumbent's initial position gives her a right to exchange with specified partners. Eligibility and priority govern claims on announced vacancies. Taking these rights as given lets us ask when a strategy-proof rule can honor them together and which rights the institution must revise when it cannot. Our analysis rests on a key assumption, the \emph{withdrawal condition}: when an incumbent receives a vacancy, her initial position is withdrawn from the round and not reassigned. This assumption is motivated by a common practice: many institutions announce a fixed list of vacancies before applications open, and applications through the vacancy channel are restricted to the positions on that list.
An incumbent still holds her initial position when the list is announced, so applicants cannot request that position through the vacancy channel in that round. If she receives a vacancy, her initial position cannot be reassigned through an exchange either, because each participant in an exchange receives another incumbent’s initial position. For instance, Spain's open and permanent competition for central-government labor staff and the internal-transfer rounds of Turkey's Ministry of Family and Social Services restrict requests to the announced posts. In Stockholm's municipal internal housing queue, homes released by successful incumbents go to the ordinary housing queue.\footnote{For Turkey, see Art.~22 of the 2013 regulation cited above and the ministry's \href{https://www.aile.gov.tr/media/201542/2025-yer-degisikligi-duyurusu.pdf}{2025 internal-transfer notice}. For Spain, see the 2023 resolution cited above, Bases~1.1 and~5.1, and the Ministry for Digital Transformation and the Civil Service, \href{https://digital.gob.es/funcion-publica/dgfp/gestiona-vida-profesional/movilidad-personas-empleadas-publicas/movilidad-personal-laboral/concurso-abierto-y-permanente-fases-finalizadas/ano-2026/concurso-abierto-y-permanente-cap-320251}{\emph{Concurso abierto y permanente CAP~1\_2026}}. For Stockholm, see the Stockholm Housing Agency's \href{https://bostad.stockholm.se/om-oss/fragor-och-svar/interna-byteskon/}{frequently asked questions on the \emph{Interna bytesk\"on}}.}

Under the withdrawal condition, exchange rights and vacancy priorities
can conflict even with two incumbents and one vacancy. For instance, suppose Ann and Bob may exchange positions and are both eligible for a vacancy, with Ann having the higher priority.
Suppose also that Ann ranks Bob's position first, Bob ranks the vacancy first, and both rank their own positions last. Keeping both in place denies them a mutually beneficial exchange. If they
exchange, the vacancy remains empty even though it is Bob's first choice. Giving the vacancy to Bob leaves Ann in her own position,
wanting the vacancy despite having higher priority. Giving the vacancy to Ann leaves Bob in his own position, and both would rather exchange their positions. Chains, such as the one giving Bob the vacancy and Ann his position, their first choices, are not permitted under the withdrawal condition, which leaves only these four assignments. Therefore, no assignment can honor their exchange rights, respect Ann's priority, and fill a vacancy that an applicant prefers to her assignment.

In this example, the conflict arises because the policy makes both Ann and Bob eligible for the vacancy. If only Ann were eligible, the conflict would disappear. Indeed, suppose that Bob prefers Ann's position to his own and that Ann ranks both Bob's position and the vacancy above her own. Giving Ann her first choice would then both honor their exchange rights and respect vacancy priorities: if she ranks Bob's position first, they exchange. If she ranks the vacancy first, she receives it and Bob keeps his position. Conflicts of this kind can thus be removed by redesigning the policy. We study such redesigns in \autoref{sec:policy}.

Even when the policy allows a strategy-proof rule respecting exchange rights and vacancy priorities, running the two channels in a fixed order, with first-stage vacancy awards or executed exchanges treated as final, need not yield such a rule.\footnote{American Airlines, for example, awards posted vacancies at crew bases by seniority and then processes mutual transfers. A vacancy transfer may not be rescinded, and a flight attendant may not rescind a mutual transfer. See its \href{https://www.apfa.org/wp-content/uploads/2024/10/CBA2024_Section22_092424.pdf}{2024 collective agreement} with the Association of Professional Flight Attendants, \S\S22.D and~22.K. France's interdepartmental procedure for primary teachers also processes transfers in a first phase and exchange cycles across departments in a second, where a teacher may improve on her first-phase assignment. See the education ministry's \href{https://www.education.gouv.fr/mutation-des-personnels-enseignants-du-premier-degre-5498}{\emph{Mutation des personnels enseignants du premier degr\'e}}.} For instance, suppose only Ann is eligible and DA allocates vacancies before TTC runs. If Ann prefers Bob's position to the vacancy, DA gives her the vacancy and withdraws her position, preventing an exchange that both prefer. This sequence also violates strategy-proofness because Ann could obtain that exchange by reporting the vacancy below her own position. Suppose instead that TTC executes exchanges before DA runs. If Ann prefers the vacancy to Bob's position, TTC nevertheless carries out the exchange, after which Bob holds Ann's initial position. Ann cannot then receive the vacancy unless the exchange is undone: the withdrawal condition bars Bob from holding the initial position of a vacancy recipient. Keeping the exchange final therefore leaves the vacancy, her first choice, unassigned. Strategy-proofness is again violated because Ann could prevent the exchange by reporting Bob's position below her own and then obtain the vacancy in the second stage. \autoref{fig:intro-example} shows the compatibility graphs and preferences in both examples.

Our \emph{propose-and-trade rule} (PTR) lets applicants pursue vacancies and exchanges simultaneously. Each applicant points to her favorite remaining option. Incumbents whose pointers form a cycle trade positions. A vacancy tentatively holds the highest-priority applicant who points to it. If awarding one vacancy withdraws a position another incumbent hoped to obtain through exchange, that incumbent may turn to a vacancy where she outranks its tentative recipient. PTR tracks these claims and finalizes mutually dependent vacancy awards together.

\begin{figure}[ht]
\centering
\begin{minipage}[t]{.38\linewidth}
\centering
\begin{tikzpicture}[every node/.style={font=\small}]
\node[draw,rectangle,minimum size=7mm] (v) at (0,1.35) {$v$};
\node[draw,circle,minimum size=10mm,inner sep=1pt] (ann) at (-1.35,0) {Ann};
\node[draw,circle,minimum size=10mm,inner sep=1pt] (bob) at (1.35,0) {Bob};
\draw[line width=1.4pt] (ann)--(bob);
\draw (ann)--(v)--(bob);
\end{tikzpicture}

\smallskip
\small
\begin{tabular}{@{}r@{\ }l@{}}
Ann: & Bob, $v$, Ann\\
Bob: & $v$, Ann, Bob\\
\mbox{} & \mbox{}
\end{tabular}

\smallskip
(a) Both eligible
\end{minipage}\hspace{.04\linewidth}
\begin{minipage}[t]{.38\linewidth}
\centering
\begin{tikzpicture}[every node/.style={font=\small}]
\node[draw,rectangle,minimum size=7mm] (v) at (0,1.35) {$v$};
\node[draw,circle,minimum size=10mm,inner sep=1pt] (ann) at (-1.35,0) {Ann};
\node[draw,circle,minimum size=10mm,inner sep=1pt] (bob) at (1.35,0) {Bob};
\draw[line width=1.4pt] (ann)--(bob);
\draw (ann)--(v);
\end{tikzpicture}

\smallskip
\small
\begin{tabular}{@{}r@{\ }l@{}}
Ann: & Bob, $v$, Ann\\
\emph{or} & $v$, Bob, Ann\\
Bob: & Ann, Bob
\end{tabular}

\smallskip
(b) Only Ann eligible
\end{minipage}

\caption{Compatibility graphs and preferences in the Ann--Bob examples.
Circles represent incumbents and squares represent vacancies. Bold
edges are exchange edges, and thin edges record eligibility. Each list ranks an incumbent's options, most
preferred first, with a name standing for that incumbent's position.
(a)~Both incumbents are eligible for $v$, where Ann has priority over Bob.
(b)~Only Ann is eligible, and the two lists show her alternative preferences.}
\label{fig:intro-example}
\end{figure}

Our approach is axiomatic. We ask which policies admit a strategy-proof,
individually rational rule that respects both exchange rights and vacancy
priorities. \emph{Individual rationality}
protects each applicant's outside option: her position for an incumbent
and remaining unmatched for a newcomer. To protect exchange rights, we impose a weak requirement, \emph{exchange-pair rationality}: it protects any permitted pairwise exchange of initial positions that would benefit one incumbent without hurting the other. To protect vacancy priorities, we impose the standard requirement, \emph{vacancy-priority respect}: if an eligible applicant prefers a vacancy to her assignment, the vacancy must go to someone with higher priority. In particular, it cannot remain empty. We refer to these two bilateral protections together as \emph{pair rationality}. We call a policy \emph{implementable}
if it admits a strategy-proof, individually rational, and pair-rational
rule on the full strict preference domain.

We show that a policy is implementable exactly when it has no \emph{conflict cycle} 
(\autoref{thm:main}). A conflict cycle is a cycle in the compatibility graph that
contains both a vacancy and an exchange edge and can be oriented so that each vacancy 
ranks its successor above its predecessor. We call a policy without a conflict cycle 
\emph{conflict-free}. Restricting protection to first-choice claims gives the same 
condition.

When PTR finalizes a group of vacancies under a conflict-free policy, any active 
applicant who outranks the tentative recipient at one of those vacancies receives 
another vacancy in the group that she prefers to it. PTR implements every such policy 
and selects the applicant-optimal core allocation at each preference profile 
(\autoref{thm:uniqueness}(i)). Every applicant weakly prefers this allocation to every 
other core allocation, and no group of incumbents can block it by trading along a 
permitted cycle.

Our results also guide how a policy must be reformed to be implementable. Changing vacancy priorities preserves eligibility and exchange permissions, so we first ask when priority reform suffices. The answer depends on \emph{exchange components}, the groups of applicants linked by chains of permitted exchanges. If each vacancy has at most one eligible applicant in every exchange component, priority reform suffices: ordering the components and letting each vacancy rank its eligible applicants in this order makes the policy conflict-free (\autoref{prop:policy}). If instead some vacancy has two eligible applicants in the same component, such as Ann and Bob in our example, no priority reform can remove the conflict. Indeed, even if vacancy priorities are ignored altogether, no strategy-proof and individually rational rule can both let compatible incumbents exchange whenever they rank each other's positions first and fill every vacancy ranked first by an eligible applicant (\autoref{prop:priority-free}). Implementability then requires revising eligibility or exchange permissions.

We next ask whether PTR is the only rule that meets our requirements. The answer depends on the exchange graph, which links incumbents through permitted exchanges. For a conflict-free policy, PTR is the only strategy-proof and individually rational rule satisfying first-choice pair rationality if and only if the exchange graph has no \emph{even hole}: a cycle through an even number of incumbents, at least four, in which no two nonconsecutive members may exchange (\autoref{thm:uniqueness}(ii)). Thus conflict cycles determine whether a rule meeting our requirements exists, while even holes determine whether institutions have a choice among such rules. When the exchange graph has an even hole, PTR is still the only strategy-proof rule that is \emph{constrained efficient}: it always selects an individually rational and pair-rational allocation that no other such allocation Pareto dominates (\autoref{cor:constrained}).

Our results also yield new characterizations of two classic rules. Without exchanges, PTR reduces to DA. Without vacancies, it reduces to TTC. The same axioms (strategy-proofness, individual rationality, and first-choice pair rationality) characterize DA when no exchanges are permitted and TTC when there are no vacancies and every exchange is permitted (\autoref{cor:boundary}). These results strengthen the characterization of DA by stability and strategy-proofness \citep{AB1994} and the characterization of TTC by strategy-proofness, individual rationality, and pair rationality \citep{EY2026}.

PTR is manipulable under every policy with a conflict cycle (\autoref{cor:robust-sp}). Under each such policy, it still protects incumbent exchanges and first-choice claims. We identify exactly which vacancy priorities it can violate. A violation can occur only when a higher-priority incumbent pursues an exchange she prefers to the vacancy, so newcomers retain full priority protection. Switching to any strategy-proof rule makes some applicant worse off than under PTR at some preference profile (\autoref{sec:robustness}).

\subsection{Related literature}\label{sec:literature}

We connect the literatures on object reallocation and priority-based
matching and contribute to the axiomatic characterization of DA and TTC. In object reallocation, \citet{Papai2003,Papai2007} studies
which restrictions on trades permit strategy-proof exchange rules when
agents may own several goods. Our
exchange graph likewise restricts feasible trades, and we ask how these
restrictions interact with eligibility and priorities at announced
vacancies. A separate question is what rights an endowment confers.
\citet{BK2019} interpret endowments as exclusion rights. Our requirement
for exchange rights, exchange-pair rationality, protects any
permitted pairwise exchange of initial positions that benefits one
incumbent without hurting the other.

Like our model, house allocation with existing tenants brings incumbents and
newcomers into one market. \citet{AS1999} introduce a strategy-proof and
individually rational rule for this setting, and \citet{SU2010}
characterize it. This rule reassigns positions released by vacancy
recipients, which our withdrawal condition rules out, and need not
respect prescribed priorities at individual vacancies.

In reallocation with priorities, rankings at the school or position
level govern the reassignment of occupied as well as vacant positions
\citep{CTT2022,CDTTU2025,Combe2023}. Ownership-adjusted DA,
studied as teacher-proposing DA by \citet{Pereyra2013} and characterized
by \citet{CS2024}, gives each incumbent first priority at her own
position. Under restrictions on ties and priority reversals,
\citet{EW2018} partially characterize weak priority structures that admit a strategy-proof
constrained-efficient rule, including structures that rank an incumbent
first at her own position and tie all other applicants there. Our model
instead separates ownership from priority: an incumbent's initial position
gives her a right to exchange, and priorities govern access to announced
vacancies.

This separation shapes the existence question. Existing acyclicity
conditions restrict priority structures
\citep{Ergin2002,Kesten2006,ErdilEhlers2010,Kojima2011,KKS2019,MT2026}.
Our condition, the absence of conflict cycles, instead restricts how
priorities interact with eligibility and permitted exchanges: a conflict
cycle links a claim to a vacancy with an exchange opportunity. Like the hierarchy condition of \citet{KK2018} under
distributional constraints, our condition concerns policies rather than
preferences.

Like PTR, local priority mechanisms \citep{RA2024} include both DA and
TTC as special cases, and trading-cycle rules built on control rights
\citep{Papai2000,PU2017} include TTC. The trading-cycle rules are Pareto efficient, so unlike
PTR they cannot reduce to DA under arbitrary priorities.

Applicant-proposing DA is strategy-proof \citep{DF1981,Roth1982MOR}, and
\citet{AB1994} characterize it by stability and one-sided
strategy-proofness. Other characterizations use monotonicity or
invariance \citep{KM2010,Morrill2013IJGT,EK2014,EK2016}, or
minimal-access stability with strategy-proofness and minimal-access
monotonicity \citep{KlausKlijn2024}. Without exchanges, individual
rationality and pair rationality together are equivalent to stability.
\autoref{cor:boundary}(i) weakens pair rationality to the first-choice
stability of \citet{DMS2018}, which tests only claims to an applicant's
first choice. With priorities fixed, strategy-proofness,
individual rationality, and this weaker requirement still
uniquely determine DA.

TTC is strategy-proof \citep{Roth1982TTC}. It is characterized by
strategy-proofness and individual rationality together with Pareto
efficiency \citep{Ma1994} or pair efficiency
\citep{Ekici2024,EkiciSethuraman2024}. \citet{EY2026} connect DA and TTC
through strategy-proofness, individual rationality, and pair
rationality, which characterize TTC in the Shapley--Scarf market. We
characterize existence in markets with both exchanges and vacancies,
and uniqueness under restricted compatibility.
\autoref{cor:boundary}(ii) strengthens their characterization by
weakening pair rationality to first-choice pair rationality. Their main
theorem instead characterizes TTC using conditions on first and second
choices. At a given profile, first-choice pair rationality and these
conditions do not imply one another.

In \autoref{thm:uniqueness}(iii), we adapt the local unanimity axiom of
\citet{HZ2025}, which requires executing every cycle formed in TTC's
first step. They characterize TTC in the Shapley--Scarf market, a
special case of our model, by strategy-proofness and local unanimity.
In that market, strategy-proofness, individual rationality, and
first-choice pair rationality together imply local unanimity and thus
characterize TTC (\autoref{cor:boundary}(ii)).

\section{Model}\label{sec:model}

Our model combines the two channels described in the introduction:
incumbents may exchange their initial positions, as in object
reallocation, and incumbents and newcomers compete for announced
vacancies under fixed priorities, as in priority-based matching.

Let $\mathcal{I}$ be a finite set of incumbents, $\mathcal{N}$ a finite
set of newcomers, and $\mathcal{V}$ a finite set of announced vacancies.
The three sets are disjoint. Each incumbent holds one 
\emph{initial position} (her endowment in the reallocation literature), which we 
denote by her label.
A newcomer holds no position. Examples include an entrant on a housing
register and a newly appointed teacher. In housing applications,
incumbents are tenants, and positions and vacancies are homes. The set of applicants is
$\mathcal{A}=\mathcal{I}\sqcup\mathcal{N}$. We use $i,j,k$ for incumbents,
$a,b$ for applicants of either kind, and $v,w$ for vacancies. Preferences
are ordinal, and there are no monetary transfers.

\subsection{Compatibility, preferences, and priorities}

The institution specifies which incumbents may receive one another's
positions and which applicants are eligible for each vacancy. We
represent these restrictions by a simple undirected \emph{compatibility
graph} $G=(\mathcal{A}\sqcup\mathcal{V},E)$ whose vertices are the
applicants and the announced vacancies. Each edge joins two incumbents
or an applicant and a vacancy. An edge $\{i,j\}\in E$ between incumbents
means that each is permitted to receive the other's position. We call
such incumbents \emph{compatible} and such an edge an \emph{exchange
edge}. An edge $\{a,v\}\in E$ means that applicant $a$ is eligible for
vacancy $v$. The \emph{exchange graph} $G[\mathcal{I}]$ is the subgraph
of $G$ induced by the incumbents. The neighborhood of a vertex $x$ in
$G$ is $N_G(x)=\{y\in\mathcal{A}\sqcup\mathcal{V}:\{x,y\}\in E\}$.

Each applicant $a$ has an \emph{outside option} $o_a$. For an incumbent
$a$, $o_a=a$ is her initial position. For a newcomer $a$, $o_a=\varnothing$
represents remaining unmatched.\footnote{We can also allow incumbents to
leave, with $\varnothing$ ranked anywhere in their preferences. An
incumbent who leaves has her initial position withdrawn. The characterizations
in \autoref{thm:main} and \autoref{thm:uniqueness} extend to this
setting, with each incumbent's option to leave treated as a vacancy for
which she alone is eligible.} Unlike a position or a vacancy,
$\varnothing$ may be received by several applicants. Applicant $a$'s set
of feasible options is
\[
O_a=\{o_a\}\cup N_G(a).
\]

Each applicant $a$ cares only about the object she receives, and her
preferences may be any strict order $P_a$ on $O_a$. We write
$x\mathrel{P_a}y$ if $a$ strictly prefers $x$ to $y$, and
$x\mathrel{R_a}y$ if $x\mathrel{P_a}y$ or $x=y$. Every combination
of these orders is admissible. An object is
\emph{acceptable} to $a$ if she ranks it above $o_a$.

Each vacancy $v$ has a prescribed strict priority order $\succ_v$ over
its eligible applicants $N_G(v)$, where $a\succ_v b$ means that $a$ has
higher priority than $b$ at $v$. Let
\mbox{$\succ=(\succ_v)_{v\in \mathcal{V}}$} denote the \emph{priority
profile}. The compatibility graph and the priority profile together form
the \emph{policy} $(G,\succ)$, which is fixed before applicants report
their preferences.

\subsection{Allocations and withdrawal}\label{subsec:allocations}
Each applicant receives exactly one object, possibly her outside option,
and each position or vacancy goes to at most one applicant. Formally,
an \emph{allocation} is a map
$\mu:\mathcal{A}\to\mathcal{I}\cup\mathcal{V}\cup\{\varnothing\}$ such
that each applicant $a$ receives an object $\mu_a\in O_a$ and no two
applicants receive the same position or vacancy. A position or vacancy received by no applicant
is unassigned.

Allocations must also satisfy the \emph{withdrawal condition}: the
initial position of any incumbent who receives a vacancy is left unassigned.
For example, if two incumbents are compatible and both are eligible for a
vacancy, they may exchange positions, or one may receive the vacancy
while the other keeps her position. A chain that instead assigns the
vacancy recipient's released position to the other incumbent violates
the withdrawal condition.

Under the withdrawal condition, every incumbent who does not receive a
vacancy keeps her initial position or exchanges it in a cycle of incumbents
(\autoref{lem:structure}). We allow simultaneous exchanges among any two or more distinct incumbents arranged in a cyclic order, in which each receives her successor’s position, provided that she is compatible with that successor.  
In \autoref{sec:pairwise}, we study limits on how many
incumbents may participate in a single exchange.

A \emph{rule} $\varphi$ is a deterministic function that selects an
allocation $\varphi(P;\succ)$ at every preference profile
$P=(P_a)_{a\in \mathcal{A}}$, and $\varphi_a(P;\succ)$ denotes applicant
$a$'s assignment. When the policy is clear from context, we simply write $\varphi(P)$ and
$\varphi_a(P)$.

\subsection{Individual and bilateral protections}

An allocation $\mu$ is \emph{individually rational (IR)} if
$\mu_a\mathrel{R_a} o_a$ for every applicant $a$.\footnote{If incumbents
may also leave, IR requires each incumbent's assignment to be weakly
preferred both to her initial position and to leaving.}

Two bilateral protections constrain allocations: one protects exchanges
between compatible incumbents, and the other protects priority at
announced vacancies.

\begin{definition}[Pair rationality]\label{def:pr}
An allocation $\mu$ satisfies \textbf{exchange-pair rationality} if no
compatible incumbents $i,j$ satisfy $j\mathrel{R_i}\mu_i$ and
$i\mathrel{R_j}\mu_j$ with at least one comparison strict. It satisfies
\textbf{vacancy-priority respect} if, whenever an eligible applicant
prefers a vacancy to her assignment, the vacancy goes to an applicant
whom it ranks above her. \textbf{Pair rationality (PR)} is the
conjunction of the two.
\end{definition}

Exchange-pair rationality applies even if $i$ already receives $j$'s
position: $i$ is then indifferent to the exchange, but the pair still
blocks if $j$ strictly prefers $i$'s position to her own
assignment.\footnote{The necessity argument in \autoref{lem:necessity} uses only claims
against incumbents who retain their initial positions when both parties
would strictly gain from exchanging. In \autoref{prop:priority-free},
SP already rules out such a claim when the claimant ranks the owner's
position first and the rule lets compatible incumbents exchange whenever
they rank each other's positions first. The uniqueness proof uses
indifference in \autoref{lem:topcycleforcing}.} Newcomers hold no exchange rights. Vacancy-priority respect implies
\emph{non-wastefulness}: no vacancy that an eligible applicant prefers
to her assignment remains unassigned.

Each notion in \autoref{def:pr} has a first-choice version, in which an
applicant can object only to being denied her first choice.

\begin{definition}[First-choice pair rationality]\label{def:fcpr}
An allocation $\mu$ satisfies \textbf{first-choice exchange-pair
rationality} if there are no compatible incumbents $i,j$ such that $i$
ranks $j$ first, $\mu_i\ne j$, and $i\mathrel{R_j}\mu_j$. It satisfies
\textbf{first-choice vacancy-priority respect} if
vacancy-priority respect holds for every vacancy $v$ and every
eligible applicant $a$ who ranks $v$ first. \textbf{First-choice pair
rationality (first-choice PR)} is the conjunction of the two.
\end{definition}

If an incumbent is denied another incumbent’s position that she ranks first, its owner must strictly prefer her assignment to the claimant’s initial position. 
PR implies first-choice PR, but the converse fails. To see this, return to the
Ann--Bob market in which both incumbents are eligible for $v$ and Ann
has higher priority (see \autoref{fig:intro-example}(a)). Giving the
vacancy to Bob and leaving Ann in her position satisfies first-choice PR.
Bob receives his first choice. Ann ranks Bob's position first, but Bob
prefers the vacancy to
Ann's position, so her first-choice exchange claim does not block this
allocation. Full PR nevertheless fails because Ann prefers the vacancy to her assignment
and has higher priority than Bob.

A rule satisfies IR, PR, or first-choice PR if the allocation it
selects satisfies the corresponding requirement at every preference
profile.

A rule $\varphi$ is \emph{strategy-proof} (SP) if, for every applicant
$a$, every profile $P$, and every strict order $P'_a$ on $O_a$,
$\varphi_a(P)\mathrel{R_a}\varphi_a(P'_a,P_{-a})$.
That is, truthful reporting is a weakly dominant strategy for every
applicant. A policy $(G,\succ)$ is \emph{implementable} if it admits an
SP, IR, and PR rule.

Without exchange edges, the model is equivalent to the
unit-capacity case of the college admissions model of \citet{GS1962},
with each applicant's outside option corresponding to remaining unmatched.
Each vacancy ranks its eligible applicants in priority order and treats
all other applicants as unacceptable. IR and PR together are equivalent
to stability. With IR imposed, first-choice PR reduces to the
first-choice stability of \citet{DMS2018}.

Without vacancies, the model is object reallocation among incumbents:
the withdrawal condition is vacuous, every newcomer remains
unmatched, and every allocation permutes the incumbents' initial positions. Under complete compatibility, this is the housing market of
\citet{SS1974}.

\section{Conflict cycles and implementation}
\label{sec:implementation}

We show that a policy is implementable exactly when it has no 
conflict cycle (\autoref{thm:main}), and we construct the propose-and-trade rule, which implements 
every such policy.

\subsection{Conflict cycles}\label{sec:no-conflict}

A \emph{cycle} in $G$ consists of at least three distinct vertices
joined consecutively by edges, with an edge also joining the
last vertex to the first. An \emph{orientation} of a cycle directs its edges
so that every vertex has exactly one incoming and one outgoing edge
within the cycle. The incoming edge comes from the vertex's
\emph{predecessor}, and the outgoing edge leads to its \emph{successor}.

Return to the Ann--Bob example in which both incumbents are eligible
for vacancy $v$ (see \autoref{fig:intro-example}(a)).
The resulting triangle has two orientations:
$\text{Ann}\to\text{Bob}\to v\to\text{Ann}$ and
$\text{Ann}\to v\to\text{Bob}\to\text{Ann}$.
In the first, $v$'s predecessor is Bob and its successor is Ann.
In the second, these roles are reversed.

\begin{definition}[Conflict cycle]\label{def:cycle}
An orientation of a cycle containing a vacancy is
\textbf{priority-consistent} if every vacancy ranks its successor above
its predecessor. A cycle is priority-consistent if it admits such an
orientation.

A \textbf{conflict cycle} is a priority-consistent cycle containing a
vacancy and an exchange edge. A policy is \textbf{conflict-free} if it
has no conflict cycle.
\end{definition}

In the triangle formed by Ann, Bob, and $v$ in \autoref{fig:intro-example}(a), Ann has
higher priority than Bob at $v$, so the first orientation
is priority-consistent. Because it contains the Ann--Bob exchange edge,
the triangle is a conflict cycle. If Bob had higher priority at $v$, the second orientation would be
priority-consistent. With both incumbents eligible, the triangle is
therefore a conflict cycle under either priority order. If only
Ann is eligible for $v$, as in \autoref{fig:intro-example}(b), the cycle
disappears and the policy is conflict-free.

We extend the market in \autoref{fig:intro-example}(b) to show that a
priority-consistent cycle need not create a conflict. Add a vacancy $w$
and two newcomers, Carol and David, who are eligible for both vacancies.
Ann and Bob may still exchange. Ann is eligible only for $v$, and Bob
is eligible for neither vacancy. Suppose
$\text{Ann}\succ_v\text{Carol}\succ_v\text{David}$ and
$\text{David}\succ_w\text{Carol}$.
The orientation $\text{Carol}\to w\to\text{David}\to v\to\text{Carol}$
is priority-consistent: at each vacancy, the successor has higher
priority than the predecessor. This cycle is the only one in the
compatibility graph, and it contains no exchange edge. The policy
is therefore conflict-free. A cycle alternates between applicants and vacancies exactly when it
contains no exchange edge. Thus a policy is conflict-free if and only if
every priority-consistent cycle containing a vacancy alternates in this way.

We also consider exchange claims by larger groups of incumbents.
A \emph{compatible incumbent cycle}
is a cyclic ordering of at least two distinct incumbents, each
compatible with her successor. It \emph{blocks} an allocation if
each member weakly prefers her successor's position to her
assignment and at least one strictly prefers it.

The \emph{core} consists of allocations that satisfy
IR and PR and admit no blocking compatible incumbent cycle. It adapts
the strong core of a housing market \citep{RP1977} to vacancies and
newcomers. To test whether a coalition can block an allocation by reallocating
its own initial positions, it suffices to check cycles. Any such blocking
reallocation decomposes into cycles, and a cycle containing a member
who strictly gains blocks on its own.

\Needspace{9\baselineskip}
\begin{theorem}[Existence]\label{thm:main}
Fix a compatibility graph $G$ and a priority profile $\succ$.
The following statements are equivalent:
\begin{enumerate}[label=(\roman*)]
\item $(G,\succ)$ is conflict-free.
\item There exists an SP, IR, and PR rule.
\item There exists an SP, IR, and first-choice PR rule.
\item There exists an SP rule selecting a core allocation at every
preference profile.
\end{enumerate}
\end{theorem}

To see why a conflict cycle prevents implementation, take a
priority-consistent orientation of it. Consider the profile at which each
applicant on the cycle ranks her successor first, her predecessor
second, and her outside option third. Every applicant outside the cycle
ranks her outside option first. At this profile, an incumbent ranks
another incumbent's position above her own only if they are neighbors
on the cycle. Because the cycle contains a vacancy, the exchange edges
along it form paths. An exchange among three or more incumbents would
require a closed loop, and these paths contain none. The withdrawal
condition rules out chains ending at a vacancy. Under IR, each applicant
on the cycle therefore either receives her outside option or is paired
with a neighbor through a swap or a vacancy assignment. IR and PR
together require every vertex of the cycle to be paired.

An odd cycle cannot be covered by such pairs, so no IR and PR
allocation exists. An even cycle admits two pairings. The exchange
edge ensures that each pairing leaves some incumbent with her second
choice. By reporting that choice below her initial position, she forces the
other pairing and obtains her first choice, contradicting
strategy-proofness. Without an exchange edge, this argument fails: in a cycle that alternates between applicants and
vacancies, all applicants prefer the same pairing, in which each receives her first choice.

The argument extends to first-choice PR because strategy-proofness supplies the missing
protection at second-choice vacancies. At the constructed profile, an
applicant on the cycle who ranks a vacancy second has higher priority there than every
other applicant who finds it acceptable. Reporting it first would therefore
secure it under IR and first-choice vacancy-priority respect, so
strategy-proofness requires her truthful assignment to be at least as
good. The same contradiction follows. The full proof is in
\appref{app:implementation}.

\subsection{The propose-and-trade algorithm}\label{sec:algorithm}

We introduce an algorithm that combines proposals to vacancies, as in DA,
with exchanges of positions, as in TTC. A vacancy can reject a lower-priority applicant
who points to it, whereas an incumbent's position remains available for trade until its owner
leaves. The difficulty is deciding when to finalize vacancy
assignments. When an incumbent's vacancy assignment is finalized, her position is
withdrawn. Another incumbent pursuing that position may then turn to a vacancy
herself. Finalizing that vacancy's assignment too early could give it to an
applicant whom this other incumbent outranks.

To handle such priority claims, the algorithm finalizes vacancy assignments in
groups. Initially, all applicants and vacancies are active, and no
vacancy has rejected anyone.
\begin{enumerate}[label=\textbf{Step \arabic*.},leftmargin=*]
\item \textbf{Pointers and rejections.} If no active applicant remains, go to
Step~4. Otherwise, each active applicant points to her favorite
among her outside option, the positions of active compatible incumbents,
and the active eligible vacancies that have not rejected her. Each vacancy
with applicants pointing to it tentatively holds the highest-priority
applicant among them and permanently rejects the others. Rejected applicants point
to their favorite remaining option. Repeat until no vacancy has more
than one applicant pointing to it.
\item \textbf{Exchanges and unmatched departures.} A \emph{pointer
cycle} consists of incumbents whose pointers form a cycle, possibly a
single incumbent pointing to her own position. If no pointer
cycle exists and no newcomer points to $\varnothing$, go to Step~3.
Otherwise, process all pointer cycles and all newcomers pointing to
$\varnothing$. Each incumbent in a pointer cycle
receives the position she points to and departs, and her position is
removed. Each newcomer pointing to $\varnothing$ departs
unmatched. Then return to Step~1.
\item \textbf{Vacancy assignments.} Since no pointer cycle exists and no
newcomer points to $\varnothing$, every active applicant's pointer path,
obtained by following her pointer and then the pointer of each incumbent
reached, ends at a vacancy. To decide which assignments to finalize
together, construct a \emph{dependency graph} whose vertices are the
vacancies that tentatively hold an applicant. It has a link $v\to w$
whenever an active applicant whose pointer path
ends at $w$ is eligible for $v$ and outranks the applicant held by $v$.
The link records a potential higher-priority claim at $v$, giving $v$
a reason to wait for the assignment at $w$ to be resolved.

A \emph{finalization group} is a maximal set of vacancies that can
reach one another by following directed links in the dependency graph,
with no link leaving the set. Such a group has no dependencies
on vacancies outside it. A vacancy in the dependency graph with no link
to another vacancy forms a group by itself. In each finalization group, finalize the
assignments together, giving each vacancy to the applicant it holds.
Remove these applicants, their initial positions (if any), and the
assigned vacancies, and return to Step~1.
\item \textbf{Completion.} The remaining vacancies stay unassigned, and the
algorithm ends.
\end{enumerate}
Each return to Step~1 follows the departure of at least one applicant, so
the algorithm reaches Step~4 after finitely many rounds.

Under a conflict-free policy, finalizing whole groups never gives a
vacancy away while an applicant with higher priority there may still
claim it. To see why, suppose
an active applicant outranks the applicant held by a vacancy $v$ in a
finalization group. She must be held by another vacancy in the
group: if she were still pursuing an exchange, her pointer path and the
links within the group would yield a conflict cycle (\autoref{lem:safe}).
Because $v$'s tentative recipient outranks every applicant $v$ has
rejected, $v$ has never rejected her and remains among her options, so
she prefers the vacancy holding her to $v$. Finalizing the group together
therefore gives her that preferred vacancy.

For each policy, the algorithm produces a unique allocation at every
preference profile. It therefore defines a rule, the
\emph{propose-and-trade rule} (PTR), denoted $\varphi^*$. Under a
conflict-free policy, PTR
selects from the core (\autoref{lem:existence}) and is strategy-proof
(\autoref{lem:sp}). Together, these lemmas prove the sufficiency part
of \autoref{thm:main}.

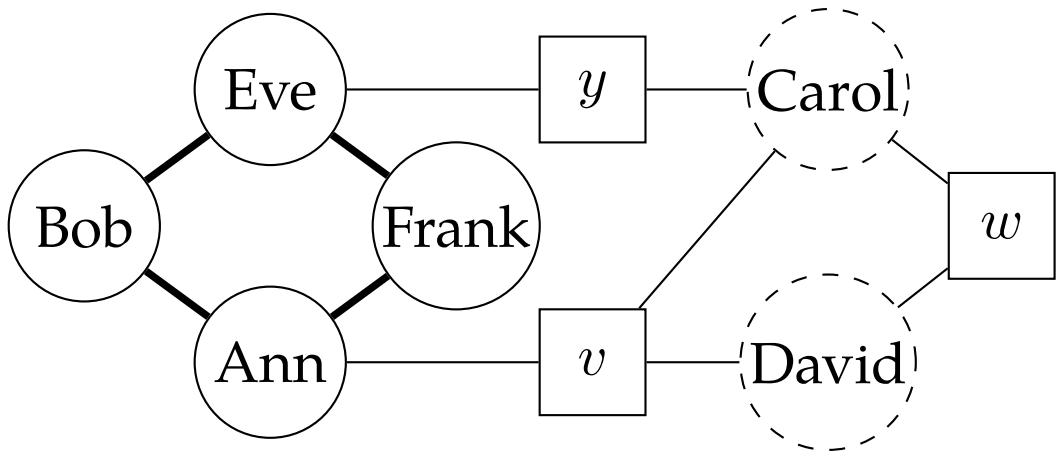
\begin{figure}[ht]
\centering
\begin{minipage}[c]{.57\linewidth}
\centering
\begin{tikzpicture}[scale=.82, every node/.style={font=\small}]
\node[draw,circle,minimum size=10mm,inner sep=1pt] (ann) at (-2,-1.1) {Ann};
\node[draw,circle,minimum size=10mm,inner sep=1pt] (bob) at (-3.5,0) {Bob};
\node[draw,circle,dashed,minimum size=10mm,inner sep=1pt] (carol) at (2.5,1.1) {Carol};
\node[draw,circle,dashed,minimum size=10mm,inner sep=1pt] (david) at (2.5,-1.1) {David};
\node[draw,circle,minimum size=10mm,inner sep=1pt] (eve) at (-2,1.1) {Eve};
\node[draw,circle,minimum size=10mm,inner sep=1pt] (frank) at (-.5,0) {Frank};
\node[draw,rectangle,minimum size=7mm] (v) at (.6,-1.1) {$v$};
\node[draw,rectangle,minimum size=7mm] (w) at (3.9,0) {$w$};
\node[draw,rectangle,minimum size=7mm] (y) at (.6,1.1) {$y$};
\draw[line width=1.4pt] (ann)--(bob)--(eve)--(frank)--(ann);
\draw (eve)--(y)--(carol)--(w)--(david)--(v)--(ann);
\draw (carol)--(v);
\end{tikzpicture}
\end{minipage}\hfill
\begin{minipage}[c]{.39\linewidth}
\centering
\small
\begin{tabular}{cl}
\toprule
Vacancy & Priority\\
\midrule
$v$ & Ann $\succ$ Carol $\succ$ David\\
$w$ & David $\succ$ Carol\\
$y$ & Eve $\succ$ Carol\\
\bottomrule
\end{tabular}
\end{minipage}
\caption{The compatibility graph and vacancy priorities in the extended
Ann--Bob example. Solid and dashed circles denote incumbents and
newcomers, respectively. Squares denote vacancies, bold edges exchanges,
and thin edges eligibility.}
\label{fig:mixed-market}
\end{figure}

To illustrate the algorithm, we extend the conflict-free market in
\autoref{sec:no-conflict} by adding incumbents Eve and Frank and
vacancy $y$. The exchange edges now form the four-cycle
Ann--Bob--Eve--Frank--Ann. Eligibility and priorities at $v$ and $w$
are unchanged. Carol and Eve are eligible for $y$, and Eve has higher
priority there. \autoref{fig:mixed-market} shows the compatibility
graph and vacancy priorities.

The policy is conflict-free. Every cycle with a vacancy and an exchange
edge uses Ann--$v$ and Eve--$y$, the only edges joining an incumbent to
a vacancy. In either orientation, $v$ or $y$ has its highest-priority
applicant as its predecessor, so neither orientation is
priority-consistent.

We run the algorithm at profile $P$ below, listing preferences from
most to least preferred and using applicants' initials as subscripts:
\[
\begin{array}{r@{\;:\;}l}
P_A&\text{Frank},\text{Bob},v,\text{Ann}\\
P_B&\text{Ann},\text{Eve},\text{Bob}\\
P_C&y,w,v,\varnothing
\end{array}
\qquad
\begin{array}{r@{\;:\;}l}
P_D&v,w,\varnothing\\
P_E&y,\text{Frank},\text{Eve},\text{Bob}\\
P_F&\text{Eve},\text{Frank},\text{Ann}.
\end{array}
\]

In Step~1, Ann points to Frank, Bob to Ann, Carol to $y$,
David to $v$, Eve to $y$, and Frank to Eve. Vacancy $y$ holds Eve
and rejects Carol, who then points to $w$. Thus $v$, $w$, and $y$ hold
David, Carol, and Eve, respectively. The incumbents' pointers form the
path $\text{Bob}\to\text{Ann}\to\text{Frank}\to\text{Eve}\to y$, so
there is no pointer cycle. Since neither newcomer points to
$\varnothing$, Step~2 sends the algorithm to Step~3.

In Step~3, Ann outranks David at $v$, and her pointer path ends at $y$,
giving the dependency link $v\to y$. Carol outranks David at $v$ and
points to $w$, giving $v\to w$. David outranks Carol at $w$ and points
to $v$, giving $w\to v$. No link leaves $y$, which holds Eve, its
highest-priority applicant. Since $\{v,w\}$ has a link to $y$, $\{y\}$
is the only finalization group. We assign $y$ to Eve and withdraw her
position. The assignments at $v$ and $w$ remain tentative. Keeping them
tentative protects Ann's priority at $v$. If Bob ranked his own
position first, he would keep it and depart in the first round. Once
Eve's position was withdrawn and Frank kept his own, Ann would point to
$v$, which would reject David.

On returning to Step~1, Frank points to his own position, which he
prefers to Ann's. Step~2 finalizes this assignment and removes Frank
and his position. At the next Step~1, Ann points to Bob, who still
points to Ann. Step~2 executes their exchange and removes both
incumbents and their positions. Carol and David continue to point to
$w$ and $v$, respectively. In the next round, Step~2 again sends the
algorithm to Step~3. Only Carol and David remain active, so the
dependency graph has just the links $v\to w$ and $w\to v$. Vacancies
$v$ and $w$ form a finalization group, so we assign $w$ to Carol and
$v$ to David. No active applicant remains, so the algorithm ends at
Step~4. In the order Ann, Bob, Carol, David, Eve, Frank,
$\varphi^*(P)=(\text{Bob},\text{Ann},w,v,y,\text{Frank})$.

The algorithm can be implemented in polynomial time
(\suppref{prop:certifying}). Under any policy, we can also check,
before finalizing a group, that every active applicant who outranks the
tentative recipient of a vacancy in the group is held by another
vacancy in that group (the conclusion of \autoref{lem:safe}). If a check fails, the
resulting procedure returns a conflict cycle. If all checks pass, it
returns a core allocation at the submitted profile. Passing all checks
at one profile does not establish that the policy is conflict-free.

\section{Optimality and uniqueness}
\label{sec:uniqueness-main}

Even under a conflict-free policy, strategy-proofness, individual
rationality, and first-choice pair rationality need not uniquely determine
PTR. Keep the compatibility graph and vacancy priorities in
\autoref{fig:mixed-market},
but change preferences. At profile $P'$, each incumbent ranks her
successor on the Ann--Bob--Eve--Frank--Ann cycle first, and both
newcomers rank remaining unmatched first.
The preferences are
\[
\begin{array}{r@{\;:\;}l}
P'_A&\text{Bob},v,\text{Ann},\text{Frank}\\
P'_B&\text{Eve},\text{Bob},\text{Ann}\\
P'_C&\varnothing,y,w,v
\end{array}
\qquad
\begin{array}{r@{\;:\;}l}
P'_D&\varnothing,v,w\\
P'_E&\text{Frank},y,\text{Eve},\text{Bob}\\
P'_F&\text{Ann},\text{Frank},\text{Eve}.
\end{array}
\]
PTR executes the four-way exchange, whereas a second rule,
constructed in the proof of \autoref{lem:holemultiplicity}, gives $v$ to Ann and $y$ to
Eve, leaving Bob and Frank in their positions. Both rules leave
Carol and David unmatched. Listing assignments in alphabetical order, we obtain
\[
\begin{aligned}
\varphi^*(P')&=(\text{Bob},\text{Eve},\varnothing,\varnothing,\text{Frank},\text{Ann}),\\
\varphi'(P')&=(v,\text{Bob},\varnothing,\varnothing,y,\text{Frank}).
\end{aligned}
\]
Both rules are SP and satisfy IR and PR at every profile. At $P'$,
however, every incumbent strictly prefers PTR's allocation, which gives
each her first choice. No compatible pair blocks the alternative
allocation, because one member of each pair ranks the other's position
below her own.

The exchange selected by PTR is a \emph{top-choice cycle}: a compatible
incumbent cycle in which every member ranks her successor's position
first. An allocation \emph{executes} a compatible incumbent cycle if
each member receives her successor's position.

The example's exchange graph is a four-cycle with no edges joining
nonconsecutive vertices. A cycle with no such edges is
\emph{induced}. An induced cycle of even length at least four is an
\emph{even hole}.

An unexecuted top-choice cycle blocks the allocation, so every core
selector executes every top-choice cycle. First-choice
exchange-pair rationality requires the execution of every
two-incumbent top-choice cycle.

\begin{theorem}[Optimality and uniqueness]
\label{thm:uniqueness}
Suppose $(G,\succ)$ is conflict-free.
\begin{enumerate}[label=(\roman*)]
\item At every profile, the core is nonempty and has a unique
applicant-optimal allocation, which every applicant weakly prefers to
every other core allocation. PTR selects this allocation and is the
unique SP core selector.
\item PTR is the only rule satisfying SP, IR, and first-choice PR
if and only if $G[\mathcal{I}]$ has no even hole. If $G[\mathcal{I}]$
has an even hole, at least two such rules satisfy PR.
\item PTR is the unique rule satisfying SP, IR, and first-choice
vacancy-priority respect that executes every top-choice incumbent
cycle at every preference profile.
\end{enumerate}
\end{theorem}

Conflict-freeness depends on the full compatibility graph, including
newcomers, and on the vacancy priorities, whereas the even-hole condition
concerns only the exchange graph. The conclusions in parts~(ii) and~(iii) also
hold under full PR, which PTR and the second rule in part~(ii) satisfy.
Part~(i) is proved in \appref{app:implementation}, and
parts~(ii)--(iii) in \appref{app:uniqueness}.

To see what applicant optimality means in part~(i), compare two core
allocations. They agree on incumbent
exchanges and on which applicants receive vacancies. Applicants whose
assignments differ move among vacancies along alternating cycles. On
each such cycle, all applicants prefer the same allocation
(\autoref{prop:comparison}). Return to profile $P$ in
\autoref{sec:algorithm}. Leaving the incumbents' assignments unchanged
and giving $v$ to Carol and $w$ to David produces another core allocation.
Both newcomers prefer their assignments under PTR. The two allocations
differ only on the alternating cycle through Carol, $v$, David, and $w$.

Strategy-proofness in part~(i) follows from applicant optimality and
agreement across core allocations on which applicants receive their
outside options. Suppose a misreport gives an applicant an object $x$
that she prefers to her truthful assignment. The manipulated allocation
stays in the core if she instead reports $x$ as her only acceptable
object, so every core allocation at that report gives her $x$. Now let
her also report as acceptable, in their true order, the objects she
truly prefers to $x$. A core allocation at this report cannot give her
the outside option, because it would then belong to the core at the
previous report. She therefore receives $x$ or an object she truly
prefers to $x$. Restoring her true ranking changes only how she ranks
objects below her assignment, so the allocation remains in the core.
This contradicts applicant optimality at her true preferences
(\autoref{lem:sp}).

For part~(iii), suppose two rules satisfy SP, IR, and first-choice
vacancy-priority respect and differ at some profile. Among the profiles where they differ, choose
one that minimizes the total number of objects applicants rank above
their outside options. The rules then agree on every
vacancy assignment, but one executes a top-choice incumbent cycle whose
members all retain their initial positions under the other rule
(\autoref{lem:min-disagreement}). Thus two distinct such rules cannot
both execute every top-choice cycle.

For part~(ii), suppose the exchange graph has no even hole. By
part~(iii), it suffices to show that every rule satisfying SP, IR, and
first-choice exchange-pair rationality executes every top-choice
incumbent cycle. This is \autoref{lem:topcycleforcing}. Its proof
first shows that these axioms force every odd top-choice cycle on any
exchange graph. Suppose a rule satisfying them leaves one unexecuted. We change one
member's report at a time. Each member whose report changes is denied her
successor's position and continues to rank it first. By SP, she cannot
obtain that position by changing the rest of her ranking, so the cycle
remains unexecuted. At the final profile, each member ranks her
successor's position first, her predecessor's second, and her own third.
Reversing the cycle would deny each member her first choice while her
successor receives the claimant's position. Under IR, if the rule
executes neither orientation, each member either swaps positions with a
neighbor or keeps her own. Because the cycle is odd, at least one member
keeps hers. Her predecessor ranks her position first, and she prefers her
predecessor's position to her own. First-choice exchange-pair rationality
therefore rules out both possibilities. Since SP excludes the original
cycle, no allocation remains.

On an even top-choice cycle of length at least four, disjoint swaps
between neighbors can cover every member without executing the cycle.
If an edge joins two
nonconsecutive members separated by an even number of edges along the
cycle, it divides the
cycle into two odd cycles. We create odd top-choice cycles at modified profiles
and use strategy-proofness to force the original even cycle
(\autoref{lem:topcycleforcing}). Every even cycle of length at least
four has such an edge exactly when the exchange graph has no even hole.
For any conflict-free policy whose exchange graph contains an even hole,
\autoref{lem:holemultiplicity} constructs a second SP, IR, and PR rule
that leaves the hole's top-choice cycle unexecuted at a profile where
PTR executes it.

Adding an efficiency requirement restores uniqueness even when the
exchange graph has an even hole. An allocation $\nu$ \emph{Pareto
dominates} $\mu$ if every applicant weakly prefers $\nu$ to $\mu$ and
some applicant strictly prefers $\nu$ to $\mu$. We call a rule \emph{constrained
efficient} if, at every profile, it
selects an IR and PR allocation that no other IR and PR allocation
Pareto dominates.

\begin{corollary}[Constrained efficiency]\label{cor:constrained}
Suppose $(G,\succ)$ is conflict-free. Then PTR is the unique
rule that is SP and constrained efficient.
\end{corollary}

Constrained efficiency excludes the second rule in the example: at
$P'$, its allocation is Pareto dominated by the four-way exchange, with
both newcomers still unmatched. Without
vacancies, the corollary characterizes TTC on any exchange graph. 
The proof is in \suppref{app:compatibility-uniqueness}.

\subsection{Characterizations of deferred acceptance and top trading cycles}

Every conflict cycle contains both a vacancy and an exchange
edge. A policy is therefore conflict-free if it has no exchange edges
or if it has no vacancies. In the first case, first-choice PR protects
claims to vacancies. In the second, it protects claims to exchanges.
Specializing \autoref{thm:uniqueness} to these two cases gives
characterizations of DA and TTC from the same axioms.

\begin{samepage}
\begin{corollary}[Common axioms for DA and TTC]\label{cor:boundary}
\leavevmode
\begin{enumerate}[label=(\roman*)]
\item If $G$ has no exchange edges, PTR is applicant-proposing
DA under $\succ$, the unique rule satisfying SP, IR, and first-choice
vacancy-priority respect.
\item If $\mathcal{V}=\varnothing$, PTR runs TTC on
$G[\mathcal{I}]$ and leaves every newcomer unmatched. It is the unique
rule satisfying SP, IR, and first-choice exchange-pair rationality if
and only if $G[\mathcal{I}]$ has no even hole. In particular, in the
Shapley--Scarf housing market, TTC is the unique rule satisfying SP, IR,
and first-choice PR.
\end{enumerate}
\end{corollary}
\end{samepage}

Without exchange edges, the core is the set of stable
allocations. By \autoref{thm:uniqueness}(i), PTR selects the
applicant-optimal stable allocation, which is the outcome of
applicant-proposing DA \citep{GS1962}. Part~(i) strengthens the
characterization of \citet{AB1994} by stability and strategy-proofness
for applicants. Under IR, it suffices to require respect for vacancy
priorities when the vacancy is the applicant's first choice.

\citet{DMS2018} introduce first-choice stability and use an
extension of it in their characterization of the \textit{immediate-acceptance
rule} (Theorem~7.3). We use first-choice stability to extend part~(i)
to schools with several seats.

\begin{remark}[DA with school capacities]\label{rem:capacities}
In the case without incumbents, we interpret applicants as
students and vacancies as schools. We allow each school to have any
fixed number of seats. Each school ranks its eligible students in
strict priority order, and each student may rank her eligible schools
and her outside option in any strict order. For an individually rational assignment,
\emph{first-choice stability} in the sense of \citeauthor*{DMS2018}
holds if and only if, whenever a student ranks a school first and does
not receive a seat there, every seat at that school goes to a student
who outranks her there. Student-proposing DA is the unique rule
satisfying SP, IR, and first-choice stability. We give the proof in 
\appref{app:uniqueness}, immediately after \autoref{lem:min-disagreement}.
\end{remark}

Without vacancies, Step~3 of PTR never applies. Each remaining
incumbent points to her favorite remaining compatible position, possibly
her own. The members of each resulting cycle trade and leave, while
newcomers leave unmatched. This is TTC on the exchange graph, attributed
to David Gale by \citet{SS1974}. Part~(ii) shows how uniqueness depends
on which exchanges are permitted. In the Shapley--Scarf market, every
pair may exchange, so the exchange graph has no even hole.
\citet{EY2026} characterize TTC in that market by SP, IR, and PR.
Part~(ii) shows that first-choice PR suffices. Their main theorem gives
a separate characterization using conditions on first and second
choices.

\needspace{8\baselineskip}
\section{Exchange permissions and priority reform}\label{sec:policy}\label{sec:pairwise}

Suppose the institution keeps vacancy eligibility and priorities fixed
and seeks to restore implementability by restricting exchanges.
Two applicants belong to the same \emph{exchange component}
exactly when a path of exchange edges in the compatibility graph $G$
connects them. Each applicant without an exchange partner, including each newcomer, forms a singleton component.

\begin{proposition}[Exchange components and implementation]\label{prop:componentcompletion}
Let $G$ and $G'$ have the same applicants, vacancies, vacancy eligibility,
and exchange components. For every priority profile $\succ$, $(G,\succ)$
is conflict-free if and only if $(G',\succ)$ is conflict-free.
\end{proposition}

To restore implementability, the institution must split at least one exchange component (\autoref{thm:main} and \autoref{prop:componentcompletion}). To see why, consider
prohibiting a direct exchange between two incumbents. If another path
of permitted exchanges still connects them, the exchange components
stay the same, and so does implementability. In \autoref{fig:intro-example}(a),
prohibiting the Ann--Bob exchange separates the two incumbents. Both
remain eligible for $v$. With no exchange edge left, either
priority order at $v$ gives a conflict-free policy.\footnote{If the institution
suspends all exchanges, PTR reduces to DA. Sri Lanka's 2025
instructions for counselling assistants suspend mutual-transfer
requests during the annual transfer process. See its Ministry of Women
and Child Affairs,
\href{https://www.childwomenmin.gov.lk/uploads/common/transfer-of-counselling-assistant---2025.pdf}{\emph{Calling applications for annual internal transfers of counselling assistants, 2025}},
27 January 2025, para.~03, item~X.} The proof of \autoref{prop:componentcompletion} is in \suppref{app:priority-free}.

Suppose instead that the institution keeps the exchange permissions and
vacancy eligibility fixed but can revise priorities. Its options again
depend on the exchange components.

\begin{corollary}[When priority reform suffices]\label{prop:policy}
Fix a compatibility graph $G$. Some priority profile admits an SP, IR,
and PR rule if and only if each vacancy has at most one eligible
applicant in each exchange component. The equivalence is unchanged
under first-choice PR.
\end{corollary}

For necessity, suppose two applicants in the same exchange component
are eligible for the same vacancy. Their eligibility edges
close an exchange path between them into a cycle with a single vacancy.
Whichever applicant the vacancy ranks higher, the cycle is
priority-consistent in the direction that leads from the vacancy to
that applicant. Thus every priority profile leaves a conflict cycle,
and \autoref{thm:main}(iii) rules out implementation even under
first-choice PR.

For sufficiency, choose an order of the exchange components and let
each vacancy rank its eligible applicants according to that order.
Since each vacancy has at most one eligible applicant from each
component, its two neighbors on a cycle must belong to different
components. A priority-consistent orientation would therefore move
from a lower-ranked component to a higher-ranked one at each vacancy.
Each exchange edge stays within one component, so the cycle could
never return to its starting component. The resulting policy is
conflict-free, so PTR satisfies SP, IR, and PR by \autoref{thm:uniqueness}(i).

Even when the condition in \autoref{prop:policy} holds, a common
seniority ranking can leave a
conflict cycle. In \suppref{fig:policy-cycle}, Ann and Bob may exchange
and newcomer Carol forms her own component. Vacancy $v$ is open to Ann
and Carol, and $w$ to Bob and Carol. If Ann outranks Carol and Carol
outranks Bob, Ann's component comes first at $v$ and Carol's at $w$.

Consider a category in which every pair of incumbents may exchange
(see \autoref{app:policy} for examples of category restrictions). If a
vacancy is open to two incumbents in the category, no priority reform can
remove their conflict. Whenever the condition in \autoref{prop:policy} fails, no SP and IR
rule can both execute swaps between compatible incumbents who rank each
other's positions first and fill every vacancy ranked first by an eligible
applicant, even when any eligible applicant may receive it
(\suppref{prop:priority-free}).

An institution may wish to keep all permitted partners while limiting
the number of incumbents who can take part in a single exchange.\footnote{American Airlines
previously matched flight attendants for mutual transfers only in
pairs. Following its 2024 collective agreement, it expanded the
procedure to exchanges involving several flight attendants while
preserving the number of flight attendants at each base. See
Association of Professional Flight Attendants,
\href{https://www.apfa.org/2024/12/09/12-09-24-contract-implementation-update-10-november-2024-recap/}{\emph{Contract Implementation Update \#10}},
9~December 2024.}

\begin{proposition}[Limits on exchange length]\label{prop:pairwise}
Fix a policy $(G,\succ)$ and an integer $L\ge 2$. There exists an SP,
IR, and first-choice PR rule that limits every exchange to at most $L$
incumbents at every preference profile if and only if the policy is
conflict-free and the exchange graph $G[\mathcal{I}]$ contains no
simple cycle with more than $L$ incumbents. When these conditions hold,
PTR meets the limit and satisfies SP, IR, and PR.
\end{proposition}

To see why the two conditions suffice, observe that every exchange
among three or more incumbents follows a simple cycle in the exchange
graph. If every such cycle has at most $L$ members, every feasible
exchange meets the limit. Conflict-freeness then ensures that PTR
satisfies SP, IR, and PR.

Conversely, a cycle with more than $L$ members rules out the limit
even if it has a \emph{chord}, an edge joining two nonconsecutive
members of the cycle. Choose an orientation of the cycle and let each
member rank her successor's position first, her predecessor's position
second, and her own position third. IR rules out trading across a
chord at this profile. As we show in \autoref{app:policy}, SP and
first-choice exchange-pair rationality force the entire cycle to trade
in the chosen direction.

In particular, a conflict-free policy can be implemented using only
bilateral exchanges exactly when the exchange graph has no cycle. In
\autoref{fig:mixed-market}, adding the Ann--Eve edge gives the
four-cycle a chord. The cycle remains, so some profile still requires
all four incumbents to exchange. Removing the Ann--Frank edge from the
original graph instead leaves a path, so PTR implements the policy using only
bilateral exchanges.

Suppose the institution starts with a conflict-free policy, wants to keep
each exchange component connected, and can withdraw permission for individual
pairs to trade. Bilateral exchanges suffice exactly when the retained edges
form a spanning tree in each component. Keeping the tree preserves
implementation by \autoref{prop:componentcompletion}. Retaining a cycle
would force an exchange of three or more incumbents at some profile by
\autoref{prop:pairwise}. If a component consists of $n$ incumbents who may
all exchange with one another, the institution retains $n-1$ of the
$n(n-1)/2$ permitted pairs.\footnote{A tree has no even hole, so PTR is also the unique SP, IR,
and first-choice PR rule under the resulting policy
(\autoref{thm:uniqueness}(ii)).}

\section{Propose-and-trade under arbitrary policies}\label{sec:robustness}

When the policy has a conflict cycle, which exchange rights and vacancy
priorities does PTR still protect?

An applicant $a$ has \emph{justified envy} of an assigned vacancy $v$ if
she is eligible for it, prefers it to her assignment, and outranks its
recipient. We call $(a,v)$ a justified-envy pair. Non-wastefulness
covers claims on unassigned vacancies.

\begin{proposition}\label{prop:robust}
Fix any policy $(G,\succ)$ and preference profile $P$, and let
$\mu=\varphi^*(P;\succ)$.
\begin{enumerate}[label=(\roman*)]
\item The allocation $\mu$ is IR, is blocked by no compatible incumbent
cycle, and leaves unassigned no vacancy that an eligible applicant
prefers to her assignment. It satisfies first-choice PR.
\item Suppose applicant $a$ has justified envy of vacancy $v$, and let
$b$ be its recipient. Then $a$ is an incumbent. Some conflict cycle has
a priority-consistent orientation in which $b,v,a$ occur consecutively.
In this orientation, the successor of $a$ is an incumbent $j$ satisfying
\[
j\mathrel{P_a}v\mathrel{P_a}\mu_a.
\]
At the step when PTR finalizes $v$, each applicant on the cycle points
to her successor, and each vacancy on the cycle goes to its predecessor.
Each applicant on the cycle weakly prefers the object she points to at
that step to her final assignment.
\end{enumerate}
\end{proposition}

By part~(i), only justified envy can keep a PTR allocation out of the
core. By part~(ii), anyone with justified envy must be an incumbent who
prefers a compatible incumbent's position to the vacancy. PTR thus
respects newcomers' priorities and each incumbent's priority at
vacancies she prefers to all her exchange options.

In \autoref{fig:intro-example}(a), Ann ranks Bob's position first and
the vacancy second. PTR gives Bob the vacancy and withdraws his
position, so Ann keeps hers. She has justified envy because she
prefers the vacancy to her own position and outranks Bob there. The
cycle in part~(ii) runs from Bob through the vacancy to Ann and back
to Bob. Now let Bob rank his own position above Ann's, keeping the
vacancy first. PTR gives the same allocation, but giving Ann the
vacancy and leaving Bob in place is now in the core. Ann outranks Bob
at the vacancy, and Bob prefers staying to exchanging. Thus PTR can
select outside a nonempty core and need not minimize justified envy.
Eliminating Ann's envy costs Bob his first choice.

\begin{proposition}
\label{thm:robust-exact-priorities}
Fix a policy $(G,\succ)$, a vacancy $v$, and eligible applicants $a,b$
with $a\succ_v b$. The following statements are equivalent:
\begin{enumerate}[label=(\roman*)]
\item At some profile $P$, $\varphi_b^*(P;\succ)=v$ and
$v\mathrel{P_a}\varphi_a^*(P;\succ)$.
\item There is a priority-consistent orientation of a conflict cycle in
which $b,v,a$ occur consecutively and the successor of $a$ is an
incumbent.
\end{enumerate}
\end{proposition}

PTR can violate $a\succ_v b$ at some profile exactly when a conflict
cycle has a priority-consistent orientation from $b$ through $v$ to
$a$ and then along an exchange edge.\footnote{\citeauthor{DE2021} study
minimal instability among efficient improvements over DA \citeyearpar{DE2021}
and among efficient, strategy-proof mechanisms, including TTC
\citeyearpar{DE2022}. \citet{OA2026} ask which priority violations can be
justified when improving on DA. \autoref{thm:robust-exact-priorities}
identifies which applicants' priority over which recipients PTR can violate
when incumbents may exchange positions.} The proof in
\autoref{app:robust-proofs} constructs a profile at which $a$ gains
by reporting $v$ first. Every conflict cycle supplies such a comparison
(proof of \autoref{cor:robust-sp}).

\begin{corollary}
\label{cor:robust-sp}
Fix a policy $(G,\succ)$. The following statements are equivalent.
\begin{enumerate}[label=(\roman*)]
\item The policy is conflict-free.
\item PTR is strategy-proof.
\item PTR respects all vacancy priorities at every preference profile.
\end{enumerate}
\end{corollary}

A newcomer whose vacancy priorities PTR always respects can still gain
by misreporting. In the example in \autoref{app:robust-proofs}, Carol
is unmatched when she reports truthfully but receives her second
choice, $v$, when she ranks it ahead of her first choice, $w$.

Could another rule improve on PTR without hurting anyone? A rule $\psi$
\emph{Pareto dominates} $\varphi$ if every applicant weakly prefers her
assignment under $\psi$ to hers under $\varphi$ at every profile and
someone strictly prefers it at some profile.

\begin{proposition}[Pareto improvements on propose-and-trade]\label{prop:ptr-undominated}
Fix any policy $(G,\succ)$. No strategy-proof rule Pareto dominates PTR.
More precisely,
\begin{enumerate}[label=(\roman*)]
\item At any profile, an allocation that Pareto dominates the PTR
allocation permutes assigned vacancies among some of their recipients,
leaves all other assignments unchanged, and creates at least one new
justified-envy pair.
\item Under any rule that Pareto dominates PTR, some applicant can gain
by lowering her outside option in her report while preserving the
relative order of all other objects.
\end{enumerate}
\end{proposition}

Without exchanges, PTR is DA, so the proposition recovers the
unit-capacity, strict-priority case of \citet*[Theorem~1]{APR2009}.\footnote{With
strict priorities, it follows from \citet{Kesten2010} that no
Pareto-efficient, strategy-proof rule Pareto dominates DA.
On the unrestricted preference domain, \citet{KestenKurino2019} show
that no strategy-proof rule Pareto dominates DA, even without exogenous
outside options. \citet{AM2019} relate Pareto dominance among IR, SP
rules to who receives an outside option.}

When a policy has a conflict cycle, every strategy-proof rule differs
from PTR because PTR is manipulable. For each conflict cycle, there are
two profiles such that every strategy-proof rule makes some applicant
worse off than under PTR at one or both. The proof of
\autoref{thm:robust-exact-priorities} constructs them. They differ only
in one incumbent's first two choices. At both, each applicant has at
most one acceptable vacancy. No recipients can gain by trading vacancies,
so both PTR allocations are Pareto efficient by \autoref{prop:ptr-undominated}(i).
A rule that makes no one worse off must reproduce both, allowing the
incumbent to gain by reversing her first two choices.

\section{Conclusion}\label{sec:chains}\label{sec:conclusion}

Under withdrawal, a strategy-proof, individually rational rule can
protect exchange rights and vacancy priorities exactly when the policy
has no conflict cycle. Propose-and-trade implements every such policy
and selects its applicant-optimal core allocation at each preference
profile. An institution facing a conflict cycle can use the exchange
components to determine whether changing priorities alone will suffice.
If it keeps the policy, propose-and-trade still protects exchange rights
and first-choice claims. Under such a policy, replacing propose-and-trade
with any strategy-proof rule makes some applicant worse off at some
preference profile.

Some institutions reassign positions released by vacancy recipients
within the same round.\footnote{Mexico's 2026--2027 rules for within-state transfers in basic education
rank applicants by seniority and add released positions before the next
applicant chooses. Spain's teacher-transfer competition includes
released posts when educational planning retains them. For Mexico, see the Unidad del Sistema para la Carrera de las Maestras y los Maestros, Secretar\'ia de Educaci\'on P\'ublica, \href{https://www.sepbcs.gob.mx/contenido/documentos/educativo/usicamm/cambiosct/normatividad/ACUERDO_CAMBIOSCT_BASICA_MS_2026_2027.pdf}{\emph{Rules for changes of workplace in basic and upper-secondary education, 2026--2027}} (2025, title translated from Spanish), Arts.~36 and~38(II). For Spain, see Royal Decree 1364/2010, Arts.~7 and~9(4).}
We call this practice \emph{reuse}. In \autoref{fig:intro-example}(a),
reuse would allow Bob to take the
vacancy and Ann to take his released position, giving both their first
choices. Now suppose released positions are also assigned by
priority, and a newcomer, Carol, wants only Bob's position and outranks
Ann there. If Bob takes the vacancy, Carol must receive his
position. Ann then keeps her own position even though she prefers the
vacancy and outranks Bob there.

Can a strategy-proof, individually rational rule protect exchange
rights and respect priorities at both announced vacancies and released
positions? Consider a benchmark with at least one announced vacancy,
unrestricted exchanges, full eligibility, and a common priority order
for vacancies and released positions. With two incumbents, such a rule
exists exactly when one of them ranks first among all applicants. With
three or more incumbents, no such rule exists (\suppref{prop:reuse}).
Which policies admit such a rule under reuse remains open in general.

\setlength{\bibhang}{0.5em} 
\bibliographystyle{ecta}
\bibliography{exchange_references}

@article{APR2009,
  author = {Abdulkadiro{\u{g}}lu, A. and Pathak, P. A. and Roth, A. E.},
  year = {2009},
  title = {Strategy-proofness versus ef\-fi\-cien\-cy in matching with indifferences: Redesigning the {NYC} high school match},
  journal = {American Economic Review},
  volume = {99},
  pages = {1954--1978},
  doi = {10.1257/aer.99.5.1954}
}

@article{AS1999,
  author = {Abdulkadiro{\u{g}}lu, A. and S{\"o}nmez, T.},
  year = {1999},
  title = {House allocation with existing tenants},
  journal = {Journal of Economic Theory},
  volume = {88},
  pages = {233--260}
}

@article{AB1994,
  author = {Alcalde, J. and Barber{\`a}, S.},
  year = {1994},
  title = {Top dominance and the possibility of strategy-proof stable solutions to matching problems},
  journal = {Economic Theory},
  volume = {4},
  pages = {417--435},
  doi = {10.1007/BF01215380}
}

@article{AM2019,
  author = {Alva, S. and Manjunath, V.},
  year = {2019},
  title = {Strategy-proof {Pareto}-improvement},
  journal = {Journal of Economic Theory},
  volume = {181},
  pages = {121--142},
  doi = {10.1016/j.jet.2019.01.004}
}

@article{Combe2023,
  author = {Combe, J.},
  year = {2023},
  title = {Reallocation with priorities and minimal envy mechanisms},
  journal = {Economic Theory},
  volume = {76},
  pages = {551--584},
  doi = {10.1007/s00199-022-01465-x}
}

@article{CS2024,
  author = {Combe, J. and Schlegel, J. C.},
  year = {2024},
  title = {Reallocation with priorities},
  journal = {Games and Economic Behavior},
  volume = {143},
  pages = {287--299},
  doi = {10.1016/j.geb.2023.11.015}
}

@article{CTT2022,
  author = {Combe, J. and Tercieux, O. and Terrier, C.},
  year = {2022},
  title = {The design of teacher assignment: Theory and evidence},
  journal = {Review of Economic Studies},
  volume = {89},
  number = {6},
  pages = {3154--3222},
  doi = {10.1093/restud/rdac002}
}

@article{Ekici2024,
  author = {Ekici, {\"O}.},
  year = {2024},
  title = {Pair-efficient reallocation of indivisible objects},
  journal = {Theoretical Economics},
  volume = {19},
  pages = {551--564},
  doi = {10.3982/TE5471}
}

@article{EkiciSethuraman2024,
  author = {Ekici, {\"O}. and Sethuraman, J.},
  year = {2024},
  title = {Characterizing the {TTC} rule via pair-efficiency: A short proof},
  journal = {Economics Letters},
  volume = {234},
  pages = {111459},
  doi = {10.1016/j.econlet.2023.111459}
}

@unpublished{EY2026,
  author = {Ekici, {\"O}. and Yenmez, M. B.},
  year = {2026},
  title = {Minimally rational reallocation of objects},
  note = {Working paper, arXiv:2609.28174, version 1},
  url = {https://arxiv.org/abs/2609.28174v1}
}

@article{Ergin2002,
  author = {Ergin, H. I.},
  year = {2002},
  title = {Efficient resource allocation on the basis of priorities},
  journal = {Econometrica},
  volume = {70},
  number = {6},
  pages = {2489--2497},
  doi = {10.1111/j.1468-0262.2002.00447.x}
}

@article{GS1962,
  author = {Gale, D. and Shapley, L. S.},
  year = {1962},
  title = {College admissions and the stability of marriage},
  journal = {American Mathematical Monthly},
  volume = {69},
  number = {1},
  pages = {9--15}
}

@article{HM2005,
  author = {Hatfield, J. W. and Milgrom, P. R.},
  year = {2005},
  title = {Matching with contracts},
  journal = {American Economic Review},
  volume = {95},
  pages = {913--935}
}

@unpublished{CDTTU2025,
  author = {Combe, J. and Dur, U. M. and Tercieux, O. and Terrier, C. and {\"U}nver, M. U.},
  year = {2025},
  title = {Market design for distributional objectives in (re)assignment: An application to improve the distribution of teachers in schools},
  note = {CEPR Discussion Paper 19893},
  url = {https://cepr.org/publications/DP19893}
}

@unpublished{MT2026,
  author = {Murra-Anton, Zeky and Thakral, Neil},
  year = {2026},
  title = {The public housing allocation problem},
  note = {Working paper, June 2026},
  url = {https://neilthakral.github.io/files/papers/match.pdf}
}

@unpublished{HZ2025,
  author = {Hu, X. and Zhang, J.},
  year = {2025},
  title = {Local unanimity in {Shapley--Scarf} housing markets},
  note = {Working paper, arXiv:2511.22027, version 1},
  url = {https://arxiv.org/abs/2511.22027v1}
}

@article{KKS2019,
  author = {Karakaya, M. and Klaus, B. and Schlegel, J. C.},
  year = {2019},
  title = {Top trading cycles, consistency, and acyclic priorities for house allocation with existing tenants},
  journal = {Journal of Economic Theory},
  volume = {184},
  pages = {104948},
  doi = {10.1016/j.jet.2019.104948}
}

@article{Ma1994,
  author = {Ma, J.},
  year = {1994},
  title = {Strategy-proofness and the strict core in a market with indivisibilities},
  journal = {International Journal of Game Theory},
  volume = {23},
  pages = {75--83},
  doi = {10.1007/BF01242849}
}

@article{Papai2000,
  author = {P{\'a}pai, S.},
  year = {2000},
  title = {Strategyproof assignment by hierarchical exchange},
  journal = {Econometrica},
  volume = {68},
  number = {6},
  pages = {1403--1433},
  doi = {10.1111/1468-0262.00166}
}

@article{PU2017,
  author = {Pycia, M. and {\"U}nver, M. U.},
  year = {2017},
  title = {Incentive compatible allocation and exchange of discrete resources},
  journal = {Theoretical Economics},
  volume = {12},
  number = {1},
  pages = {287--329},
  doi = {10.3982/TE2201}
}

@unpublished{RA2024,
  author = {Root, J. and Ahn, D. S.},
  year = {2024},
  title = {Local priority mechanisms},
  note = {Working paper, arXiv:2309.04020, version 2},
  url = {https://arxiv.org/abs/2309.04020v2}
}

@article{Roth1982TTC,
  author = {Roth, A. E.},
  year = {1982},
  title = {Incentive compatibility in a market with indivisible goods},
  journal = {Economics Letters},
  volume = {9},
  number = {2},
  pages = {127--132},
  doi = {10.1016/0165-1765(82)90003-9}
}

@article{RP1977,
  author = {Roth, A. E. and Postlewaite, A.},
  year = {1977},
  title = {Weak versus strong domination in a market with indivisible goods},
  journal = {Journal of Mathematical Economics},
  volume = {4},
  number = {2},
  pages = {131--137},
  doi = {10.1016/0304-4068(77)90004-0}
}

@article{Sethuraman2016,
  author = {Sethuraman, J.},
  year = {2016},
  title = {An alternative proof of a characterization of the {TTC} mechanism},
  journal = {Operations Research Letters},
  volume = {44},
  number = {1},
  pages = {107--108},
  doi = {10.1016/j.orl.2015.11.010}
}

@article{SS1974,
  author = {Shapley, L. S. and Scarf, H.},
  year = {1974},
  title = {On cores and indivisibility},
  journal = {Journal of Mathematical Economics},
  volume = {1},
  number = {1},
  pages = {23--37},
  doi = {10.1016/0304-4068(74)90033-0}
}

@article{SU2010,
  author = {S{\"o}nmez, T. and {\"U}nver, M. U.},
  year = {2010},
  title = {House allocation with existing tenants: A characterization},
  journal = {Games and Economic Behavior},
  volume = {69},
  number = {2},
  pages = {425--445}
}

@article{KK2018,
  author = {Kamada, Y. and Kojima, F.},
  year = {2018},
  title = {Stability and strategy-proofness for matching with constraints: A necessary and sufficient condition},
  journal = {Theoretical Economics},
  volume = {13},
  number = {2},
  pages = {761--793},
  doi = {10.3982/TE2717}
}

@article{KM2010,
  author = {Kojima, F. and Manea, M.},
  year = {2010},
  title = {Axioms for deferred acceptance},
  journal = {Econometrica},
  volume = {78},
  number = {2},
  pages = {633--653},
  doi = {10.3982/ECTA7443}
}

@inproceedings{DMS2018,
  author = {Dur, U. M. and Mennle, T. and Seuken, S.},
  year = {2018},
  title = {First-choice maximal and first-choice stable school choice mechanisms},
  booktitle = {Proceedings of the 2018 ACM Conference on Economics and Computation},
  publisher = {ACM},
  pages = {251--268},
  doi = {10.1145/3219166.3219201}
}

@article{Morrill2013IJGT,
  author = {Morrill, T.},
  year = {2013},
  title = {An alternative characterization of the deferred acceptance algorithm},
  journal = {International Journal of Game Theory},
  volume = {42},
  pages = {19--28}
}

@article{DF1981,
  author = {Dubins, L. E. and Freedman, D. A.},
  year = {1981},
  title = {Machiavelli and the {Gale--Shapley} algorithm},
  journal = {American Mathematical Monthly},
  volume = {88},
  pages = {485--494}
}

@article{Roth1982MOR,
  author = {Roth, A. E.},
  year = {1982},
  title = {The economics of matching: Stability and incentives},
  journal = {Mathematics of Operations Research},
  volume = {7},
  pages = {617--628}
}

@article{Pereyra2013,
  author = {Pereyra, J. S.},
  year = {2013},
  title = {A dynamic school choice model},
  journal = {Games and Economic Behavior},
  volume = {80},
  pages = {100--114}
}

@article{BK2019,
  author = {Balbuzanov, I. and Kotowski, M. H.},
  year = {2019},
  title = {Endowments, exclusion, and exchange},
  journal = {Econometrica},
  volume = {87},
  pages = {1663--1692}
}

@article{EK2014,
  author  = {Ehlers, L. and Klaus, B.},
  year    = {2014},
  title   = {Strategy-proofness makes the difference: Deferred-acceptance with responsive priorities},
  journal = {Mathematics of Operations Research},
  volume  = {39},
  pages   = {949--966},
  doi     = {10.1287/moor.2014.0662}
}

@article{EK2016,
  author  = {Ehlers, L. and Klaus, B.},
  year    = {2016},
  title   = {Object allocation via deferred-acceptance: Strategy-proofness and comparative statics},
  journal = {Games and Economic Behavior},
  volume  = {97},
  pages   = {128--146},
  doi     = {10.1016/j.geb.2016.04.001}
}

@article{KlausKlijn2024,
  author  = {Klaus, B. and Klijn, F.},
  year    = {2024},
  title   = {Minimal-access rights in school choice and the deferred acceptance mechanism},
  journal = {Mathematics of Operations Research},
  volume  = {49},
  number  = {3},
  pages   = {1487--1501},
  doi     = {10.1287/moor.2022.0275}
}

@article{Papai2003,
  author = {P{\'a}pai, S.},
  year = {2003},
  title = {Strategyproof exchange of indivisible goods},
  journal = {Journal of Mathematical Economics},
  volume = {39},
  pages = {931--959},
  doi = {10.1016/S0304-4068(03)00053-3}
}

@article{Papai2007,
  author = {P{\'a}pai, S.},
  year = {2007},
  title = {Exchange in a general market with indivisible goods},
  journal = {Journal of Economic Theory},
  volume = {132},
  pages = {208--235},
  doi = {10.1016/j.jet.2005.05.012}
}

@article{EW2018,
  author = {Ehlers, L. and Westkamp, A.},
  year = {2018},
  title = {Strategy-proof tie-breaking in matching with priorities},
  journal = {Theoretical Economics},
  volume = {13},
  pages = {1009--1042},
  doi = {10.3982/TE2547}
}

@article{ErdilEhlers2010,
  author = {Ehlers, L. and Erdil, A.},
  year = {2010},
  title = {Efficient assignment respecting priorities},
  journal = {Journal of Economic Theory},
  volume = {145},
  pages = {1269--1282},
  doi = {10.1016/j.jet.2010.02.007}
}

@article{Kojima2011,
  author = {Kojima, F.},
  year = {2011},
  title = {Robust stability in matching markets},
  journal = {Theoretical Economics},
  volume = {6},
  pages = {257--267},
  doi = {10.3982/TE780}
}

@article{DE2021,
  author = {Do{\u{g}}an, B. and Ehlers, L.},
  year = {2021},
  title = {Minimally unstable {Pareto} improvements over deferred acceptance},
  journal = {Theoretical Economics},
  volume = {16},
  pages = {1249--1279},
  doi = {10.3982/TE4257}
}

@article{DE2022,
  author = {Do{\u{g}}an, B. and Ehlers, L.},
  year = {2022},
  title = {Robust minimal instability of the top trading cycles mechanism},
  journal = {American Economic Journal: Microeconomics},
  volume = {14},
  number = {4},
  pages = {556--582},
  doi = {10.1257/mic.20200148}
}

@unpublished{OA2026,
  author = {Ortega, J. and Arribillaga, R. P.},
  year = {2026},
  title = {Justifiable priority violations},
  note = {Working paper, arXiv:2604.06396},
  url = {https://arxiv.org/abs/2604.06396}
}

@article{Kesten2010,
  author = {Kesten, O.},
  year = {2010},
  title = {School choice with consent},
  journal = {Quarterly Journal of Economics},
  volume = {125},
  number = {3},
  pages = {1297--1348},
  doi = {10.1162/qjec.2010.125.3.1297}
}

@article{KestenKurino2019,
  author = {Kesten, O. and Kurino, M.},
  year = {2019},
  title = {Strategy-proof improvements upon deferred acceptance: A maximal domain for possibility},
  journal = {Games and Economic Behavior},
  volume = {117},
  pages = {120--143},
  doi = {10.1016/j.geb.2019.05.010}
}

@article{Kesten2006,
  author = {Kesten, O.},
  year = {2006},
  title = {On two competing mechanisms for priority-based allocation problems},
  journal = {Journal of Economic Theory},
  volume = {127},
  number = {1},
  pages = {155--171},
  doi = {10.1016/j.jet.2004.11.001}
}

\appendix
\numberwithin{theorem}{section}
\numberwithin{lemma}{section}
\numberwithin{proposition}{section}
\numberwithin{corollary}{section}
\numberwithin{example}{section}
\numberwithin{definition}{section}
\numberwithin{remark}{section}
\section{Implementation proofs}\label{app:implementation}
\begin{lemma}\label{lem:structure}
Let $\mu$ be an allocation, and let $S$ be the set of incumbents who do
not receive a vacancy. Then $\mu$ maps $S$ bijectively onto the
positions of $S$, so every incumbent in $S$ keeps her initial position or
belongs to a cycle of compatible incumbents in $S$, each receiving her
successor's position. Every incumbent outside $S$ and every matched
newcomer receives a distinct eligible vacancy. The positions of
incumbents outside $S$ are unassigned. Conversely, every assignment of
this form is an allocation.
\end{lemma}

\begin{proof}
\looseness=-1\relax For $i\in S$, $\mu_i$ is an incumbent's position. Its owner lies in
$S$, since the withdrawal condition excludes positions of vacancy recipients. Thus
$\mu$ maps the finite set $S$ injectively into itself and hence
bijectively. The bijection decomposes into retentions and exchange
cycles, whose edges are compatible by feasibility. Every other matched
applicant receives a distinct vacancy, and the withdrawal condition leaves vacancy
recipients' positions unassigned. Conversely, an assignment of this
form is compatible, injective on positions and vacancies, and satisfies
the withdrawal condition.
\end{proof}

\subsection{Allocation permutations}\label{app:slots}

For an allocation $\alpha$ and a vacancy $v$ assigned under $\alpha$,
write $\alpha^{-1}(v)$ for its recipient. Let $M_\alpha$
consist of all incumbents, the newcomers matched under $\alpha$, and the
vacancies assigned under $\alpha$, and define $\sigma_\alpha$ on
$M_\alpha$ by $\sigma_\alpha(a)=\alpha_a$ for an applicant $a\in
M_\alpha$ and $\sigma_\alpha(v)=\alpha^{-1}(v)$ for a vacancy $v\in
M_\alpha$. Thus the map sends an applicant to the position or vacancy she
receives and an assigned vacancy to its recipient. Unmatched newcomers
and unassigned vacancies are left out.

\begin{lemma}\label{lem:assignment-permutation}
For every allocation $\alpha$, $\sigma_\alpha$ is a permutation of
$M_\alpha$.
\end{lemma}

\begin{proof}
By \autoref{lem:structure}, $\sigma_\alpha$ consists of disjoint
retentions and exchange cycles of incumbents and two-element cycles
formed by each vacancy recipient and her vacancy. These cycles cover
$M_\alpha$, so $\sigma_\alpha$ is a permutation.
\end{proof}

In the graph-matching arguments, we call an applicant \emph{unmatched}
under $\alpha$ when she receives her outside option, including an
incumbent who retains her position. A vacancy is unmatched when it is
unassigned.

\subsection{Necessity: a conflict cycle defeats implementation}\label{sec:necessity}

\begin{lemma}\label{lem:necessity}
If the policy has a conflict cycle, no SP, IR, and first-choice PR
rule exists. In particular, no SP, IR, and PR rule exists.
\end{lemma}
\begin{proof}
Choose a conflict cycle $\Gamma=(c_0,\ldots,c_{m-1})$, oriented so that
it is priority-consistent. Let $P$ be the profile at which each
applicant on $\Gamma$ ranks her successor first, her predecessor second,
and her outside option third, with all other objects below her outside
option, and each applicant off $\Gamma$ ranks her outside option first.
Vacancy priorities are unchanged. Both neighbors of a newcomer on $\Gamma$ are vacancies.

Under IR, every applicant receives her outside option or an object
adjacent to her on $\Gamma$. By \autoref{lem:structure}, an allocation
consists of retentions, exchange cycles of incumbents, applicant--vacancy
pairs, and unmatched newcomers, and an exchange cycle of three or more
incumbents is impossible, because the incumbent edges of $\Gamma$ form a
proper subset of the edges of a cycle. Hence every applicant who
does not receive her outside option forms a pair with a neighbor on
$\Gamma$: two adjacent incumbents exchange, or an applicant receives an
adjacent vacancy.

Let $\varphi$ be an SP, IR, and first-choice PR rule. We first show that
$\varphi(P)$ matches every vertex of $\Gamma$. The same argument applies
to the modified profile at the end of this proof. Suppose a vertex $c$
is unmatched, and let $p$ be its predecessor on $\Gamma$.
\begin{enumerate}[label=(\alph*)]
\item If $c$ is a vacancy, then $p$ is an applicant, who ranks $c$ first,
and $c$ is unassigned, violating first-choice vacancy-priority respect.
\item If $c$ and $p$ are applicants, they are incumbents. $p$ ranks
position $c$ first and does not receive it, while $c$ keeps her
initial position and ranks position $p$ above it, violating first-choice
exchange-pair rationality.
\item If $c$ is an applicant and $p$ is a vacancy, let $c$ instead
report $p$ first and her outside option second, with all other objects
below. Then the only applicants who rank $p$ above their outside options
are $c$ and the predecessor of $p$ on $\Gamma$, whom $p$ ranks below $c$
by priority-consistency, so IR and first-choice vacancy-priority respect
give $p$ to $c$. By SP at $P$, $c$ receives an object she weakly prefers
to $p$, so she is not unmatched.
\end{enumerate}
Hence $\varphi(P)$ is a perfect matching of $\Gamma$ into adjacent
pairs. Only case~(c) needs strategy-proofness.

For odd $m$, no perfect matching exists, a contradiction. For even $m$,
there are exactly two perfect matchings, the two alternating edge sets.
Both satisfy IR and PR, hence first-choice PR: every applicant receives the object at
the other end of her pair, which she ranks above her outside option. The
later endpoint of an edge of $\Gamma$ outside the matching is paired
with its own successor, so it either receives its first choice or, if a
vacancy, is assigned to an applicant outranking its predecessor by
priority-consistency. An edge outside $\Gamma$ has an applicant endpoint
who ranks the object across it below her outside option. For a pair that exchanges initial positions, both comparisons are equalities.

In either matching, one parity class is matched to predecessors.
An exchange edge joins the two classes, so some incumbent $i$ receives
the object at her predecessor on $\Gamma$, her second choice.
Let her report her successor first and her initial position second, with all
other objects below. IR then removes only her predecessor edge, leaving
a path with an even number of vertices. Cases~(a)--(c) exclude unmatched
vertices other than $i$, because $i$ still ranks her successor first and
no other report has changed. But a matching on this path leaves an even
number of vertices unmatched, so $i$ must also be matched. The path's unique
perfect matching is the other alternating matching. Incumbent $i$
therefore gains her true first choice.
\end{proof}

\subsection{Feasibility and core selection}
The finalization groups in Step~3 are the sink strongly connected
components of the dependency graph. We call them \emph{sink components}
in the proofs.

At Step~3, which applies only when Step~2 has nothing to do, every
newcomer points to a vacancy and the pointers form no cycle of
incumbents. Let $r(a)$ denote the vacancy
at the end of active applicant $a$'s pointer path and $h(v)$ the applicant
held by vacancy $v$, its tentative recipient.

\begin{lemma}\label{lem:safe}
Suppose the policy is conflict-free. If $v$ lies in a sink component
finalized in Step~3, every active applicant whom $v$ ranks above $h(v)$
is herself held by a vacancy
in that same component.
\end{lemma}
\begin{proof}
We expand the dependency graph to include the active applicants
and the vacancies that tentatively hold applicants. Each applicant has an
arc along her current pointer, to the owner of the position or to the
vacancy she points to. Each such vacancy $v$ has an arc to every
active eligible applicant ranked above $h(v)$.

No applicant points to a newcomer, so a directed cycle of applicants
would consist of incumbents. No such cycle remains at this step.
A directed cycle through a vacancy
enters it from the applicant it holds, the only one pointing to it, and leaves
it toward an applicant of strictly higher priority. Since each arc joins
vertices adjacent in $G$, this is a priority-consistent cycle of $G$
containing a vacancy. Because the policy is conflict-free, it
contains no incumbent--incumbent arc.

\looseness=-1\relax Now let $v$ lie in a sink component and let $a$ be an active eligible
applicant whom $v$ ranks above $h(v)$. The link $v\to r(a)$ stays
within that component, so $r(a)$ can reach $v$. Replacing dependency
links by their applicant pointer paths gives both the path
$v\to a\to\cdots\to r(a)$ and a return path to $v$ in the expanded
dependency graph. Their vertices lie in one strongly connected component, so
every arc on the first path lies on a directed cycle.
No such arc joins two incumbents. Hence $a$
points directly to the vacancy $r(a)$ and, as the only applicant
pointing to it, is held by it, proving the claim.
\end{proof}

We write $\Acore(P;\succ)$ for the core at preference profile $P$ under
priorities $\succ$.

\begin{lemma}\label{lem:existence}
Suppose the policy is conflict-free. At every preference profile $P$,
the algorithm terminates at an allocation in $\Acore(P;\succ)$.
\end{lemma}
\begin{proof}
Each rejection permanently removes an applicant--vacancy option. If
applicants remain after rejections stop and Step~2 does not apply,
their pointer paths end at vacancies holding applicants. The dependency
graph is therefore finite and nonempty, so it has a sink component
and Step~3 applies. Every
operation of Steps~2 and~3 removes at least one applicant, so the
algorithm terminates.

Step~2 removes the members and positions of each executed cycle.
Unmatched newcomers remove no position. Step~3 removes each assigned
vacancy, its recipient, and her position, if any. Each assignment is
feasible, no object is assigned twice, and vacancy recipients' positions
are withdrawn, so the output is an allocation. IR holds because each
applicant receives her favorite available option and could still choose
her outside option at departure.

An applicant held by an active vacancy keeps pointing to it until she is
rejected in favor of a higher-priority applicant or the pair is finalized. Her
options only shrink, and an applicant pointing to a vacancy cannot
depart in Step~2. A vacancy that holds an applicant therefore stays occupied until
finalization, and each new tentative recipient has higher priority
than the previous one. Its eventual recipient outranks every applicant it rejected.
A vacancy left unassigned never held or rejected anyone.

Suppose a compatible incumbent cycle blocks the output $\mu$. Let $i$ be
a member who departs no later than any other member, and let $j$ be her
successor. Position $j$ is still active when $i$ departs, positions
never reject, and $i$ receives her favorite available object, so the
weak improvement forces $\mu_i=j$, and she departs in an exchange cycle
together with $j$. Repeating the argument along the blocking cycle, all
comparisons are equalities, contradicting the strict improvement. In
particular, no compatible pair of incumbents blocks.

Suppose $v\mathrel{P_a}\mu_a$ for an eligible applicant $a$. If $v$
rejected $a$, its eventual recipient outranks her. Otherwise, $a$
cannot depart before or with $v$: she would receive an object weakly
preferred to the available vacancy. Since $a$ eventually departs,
$v$ is assigned before her. This also proves non-wastefulness.
By \autoref{lem:safe}, every active applicant whom $v$ ranks above its
tentative recipient departs with $v$. Since $a$ remains active,
$v$'s recipient outranks her. This proves vacancy-priority respect
and PR. Only this use of \autoref{lem:safe} requires the policy to be
conflict-free.
\end{proof}

\subsection{Comparing core allocations}\label{sec:comparison}

\begin{lemma}\label{prop:comparison}
Suppose $(G,\succ)$ is conflict-free, and let $\mu,\mu'\in\Acore(P;\succ)$.
\begin{enumerate}[label=(\roman*)]
\item The same incumbents retain their initial positions in both allocations,
and $\mu_i=k$ if and only if $\mu'_i=k$ for all incumbents $i$ and $k$.
\item The same newcomers remain unmatched in both allocations, and the
same vacancies are assigned in both.
\item The applicants and vacancies whose assignments differ form disjoint
cycles $(a_1,v_1,\ldots,a_m,v_m)$, with $m\ge2$ and indices read
cyclically, on which one allocation gives $v_t$ to $a_t$ and the other
gives $v_t$ to $a_{t+1}$ for every $t$.
\item On each such cycle, all applicants strictly prefer the same
allocation, and every vacancy gives higher priority to its recipient
under the other allocation.
\end{enumerate}
\end{lemma}

\begin{proof}
Fix $P$ and $\mu,\mu'\in\Acore(P;\succ)$, and let
$U=M_\mu\cup M_{\mu'}$. For an applicant $a\in U$, let $f(a)$ be
her preferred object among $\mu_a$ and $\mu'_a$. For a vacancy
$v\in U$, let $f(v)$ be its higher-priority recipient under the two
allocations, ignoring an allocation under which it is unassigned.
IR ensures that a matched newcomer prefers her vacancy to
$\varnothing$, so every $f(x)$ is defined and equals
$\sigma_\alpha(x)$ for some $\alpha\in\{\mu,\mu'\}$ with
$x\in M_\alpha$. Thus $f$ maps $U$ into itself.

Consider a cycle $C$ of $f$. If $C$ contains no vacancy, every member
is an incumbent, since no applicant points to a newcomer. A singleton
keeps her initial position under both allocations by IR. On a longer cycle,
each member weakly prefers her successor's position to both assignments,
so the cycle blocks either allocation unless both execute it.

If $C=(a,v)$, then $\mu_a=\mu'_a=v$. Otherwise, say
$\mu_a=v\ne\mu'_a$, applicant $a$ prefers $v$ to $\mu'_a$, while
$f(v)=a$ means that $v$ is unassigned under $\mu'$ or assigned below
$a$ in priority, violating vacancy-priority respect.

If $C$ contains a vacancy and has length at least three, each vacancy's
predecessor and successor are distinct recipients under the two
allocations, and it ranks its successor higher. Hence $C$ is a
priority-consistent cycle containing a vacancy. Because the policy
is conflict-free, $C$ has no exchange edges, so $C=(a_1,v_1,\ldots,a_m,v_m)$,
with $m\ge2$. Relabel the allocations so that $\mu_{a_1}=v_1$. Then
$\mu'_{a_2}=v_1$, and $f(a_2)=v_2$ forces $\mu_{a_2}=v_2$.
Continuing around $C$ gives $\mu_{a_t}=v_t$ and
$\mu'_{a_{t+1}}=v_t$ for every $t$. All applicants on $C$ strictly
prefer $\mu$, and all its vacancies rank their $\mu'$-recipients higher.

In each case, $C\subseteq M_\mu\cap M_{\mu'}$ and
$\sigma_\mu(C)=C=\sigma_{\mu'}(C)$. Every vertex of $U$ must lie
on such a cycle. Otherwise a path of $f$ first enters a cycle $C$ at
$y=f(z)$ with $z\notin C$. Choose $\alpha\in\{\mu,\mu'\}$ with
$y=\sigma_\alpha(z)$. Since $\sigma_\alpha(C)=C$, some $w\in C$
also satisfies $\sigma_\alpha(w)=y$, contradicting injectivity
(\autoref{lem:assignment-permutation}).

The allocations agree on incumbent positions, as every applicant on a
cycle with a vacancy receives one under both. Every newcomer matched or
vacancy assigned under either lies on such a cycle. The allocations agree on the
two-member cycles and differ exactly on the longer alternating cycles.
\end{proof}

\subsection{Strategy-proofness and characterization of the rule}\label{sec:sp}

Throughout the remainder of this appendix, assume the policy is
conflict-free.

\begin{lemma}\label{lem:algorithmoptimal}
At every preference profile $P$, $\varphi^*(P;\succ)$ is the unique
applicant-optimal member of $\Acore(P;\succ)$.
\end{lemma}
\begin{proof}
Let $\mu=\varphi^*(P;\succ)$.
It belongs to $\Acore(P;\succ)$ by \autoref{lem:existence}. Let $\mu'$ be any other core allocation, and set $\Delta=\{a\in \mathcal{A}:\mu'_a\mathrel{P_a}\mu_a\}$. By
\autoref{prop:comparison}(iii) and~(iv), members of $\Delta$ receive
vacancies in both allocations. All applicants on an alternating cycle
of disagreement prefer the same allocation. Hence, whenever
$a\in\Delta$, the applicant who receives $\mu'_a$ under $\mu$ also
belongs to $\Delta$. Suppose that $\Delta\ne\varnothing$. During the execution,
an applicant's pointer moves only to less-preferred objects, because her
options shrink only through rejection or removal. Consider the first
event of either kind: an applicant $a\in\Delta$ is rejected by $\mu'_a$,
or a member of $\Delta$ departs. Rejections within a batch can be
ordered arbitrarily, since each rejected applicant chose her pointer
before the batch.

If the first event is the rejection of $a$ by $v=\mu'_a$, let $b$ be the
applicant that $v$ holds. Then $b\succ_v a$, and vacancy-priority respect
for $\mu'$ implies $\mu'_b\mathrel{P_b} v$, since $v$ goes to $a$ under
$\mu'$. Because $b$ currently points to $v$, $v\mathrel{R_b}\mu_b$, so
$b\in\Delta$. Since she prefers $\mu'_b$ to $v$, that vacancy was
unavailable when she chose $v$: it had rejected her, or it had been
removed together with its recipient under $\mu$, who also lies in
$\Delta$. Thus a member of $\Delta$ was rejected by her
$\mu'$-assignment or departed earlier, contradicting our choice of
the first event.

If the first event is the departure of some $a\in\Delta$, she points to $\mu_a$, which she
ranks below $\mu'_a$. Thus the vacancy $\mu'_a$ had already rejected her
or departed with its recipient under $\mu$, who is in $\Delta$.
Either event contradicts the choice of the first event.

Objects finalized together are still available when pointers are
chosen, so the argument also covers simultaneous departures.
Thus $\Delta$ is empty. Every applicant weakly prefers the PTR
allocation to every other core allocation, and strict preferences
give uniqueness.
\end{proof}

\begin{lemma}\label{lem:sp}
The unique SP core selector is $\varphi^*$. Under every other core selector,
some applicant gains by raising her outside option while preserving the
order of all other objects.
\end{lemma}
Following \citet[Theorems~10--11]{HM2005}, we first make a single object
acceptable. We then also make every object truly preferred to it acceptable,
preserving their true order.

\begin{proof}
By \autoref{lem:algorithmoptimal}, $\varphi^*(P;\succ)$ is the
applicant-optimal core allocation at every profile. All core allocations
at a given profile agree on whether $a$ receives $o_a$: for an incumbent,
by \autoref{prop:comparison}(i), and for a newcomer, by
\autoref{prop:comparison}(ii).

Suppose applicant $a$ can profitably misreport $P'_a$ at the true
profile $P$, obtaining
$x=\varphi_a^*(P'_a,P_{-a};\succ)\mathrel{P_a}\varphi_a^*(P;\succ)$.
IR at $P$ gives $x\mathrel{P_a}o_a$. Consider two reports:
\begin{enumerate}[label=(\alph*)]
\item $S_a$: put $x$ first, $o_a$ second, and every other object below
$o_a$.
\item $T_a$: preserve the true order of all objects weakly above $x$,
then put $o_a$, then all remaining objects.
\end{enumerate}
The allocation $\varphi^*(P'_a,P_{-a};\succ)$ remains in
$\Acore(S_a,P_{-a};\succ)$, where it gives $a$ her first choice $x$: a
block involving $a$ would have to give her $x$ again, an equality already
present at $P'_a$, and all other comparisons are unchanged. All
allocations in $\Acore(S_a,P_{-a};\succ)$ therefore give $a$ an object
other than $o_a$, and since $x$ is her only other acceptable object
under $S_a$, IR implies that each of them gives her $x$.

Choose $\tau\in\Acore(T_a,P_{-a};\succ)$, which is nonempty by
\autoref{lem:existence}. If $\tau$ gave $a$ her outside option, it would
also belong to $\Acore(S_a,P_{-a};\succ)$, because shrinking the set of
objects that an applicant ranks above her assignment $o_a$ cannot create
a block. But every core allocation at $(S_a,P_{-a})$ gives $a$ the object $x$.

Hence $\tau_a\mathrel{R_a}x$ in the true order. Restoring $P_a$ keeps
$\tau$ in the core because every object restored above $o_a$ lies below
$x$ and hence below $\tau_a$. Since $a$ prefers $x$ to her PTR
assignment at $P$, this contradicts applicant optimality and proves SP.

For uniqueness, suppose a core selector $\varphi$ differs from PTR at
$P$. By applicant optimality and IR, some applicant $a$ prefers
$x=\varphi^*_a(P;\succ)$ to $\varphi_a(P;\succ)$ and ranks $x$ above
$o_a$. Form $Q$ by placing $o_a$ immediately below $x$, preserving
the order of all other objects. The PTR allocation at $P$ remains in
the core at $Q$ because $a$'s comparisons with $x$ and all other
applicants' preferences are unchanged. Since this allocation gives
$a$ the object $x$, \autoref{prop:comparison} rules out $o_a$ for
her in every core allocation at $Q$. By IR, $\varphi_a(Q;\succ)$ is
$x$ or an object she prefers to $x$. Thus $a$ gains by raising $o_a$,
a deviation that SP rules out under PTR.
\end{proof}

\begin{proof}[Proofs of \autoref{thm:main} and \autoref{thm:uniqueness}(i)]
For \autoref{thm:main}, \autoref{lem:necessity} shows that (iii) implies (i). Under (i),
\autoref{lem:existence} gives a core allocation, and
\autoref{lem:algorithmoptimal} shows that PTR selects the unique
applicant-optimal core allocation.
Then \autoref{lem:sp} gives (iv) and the uniqueness of
$\varphi^*$ among SP core selectors. These conclusions also prove
\autoref{thm:uniqueness}(i). For the remaining implications in
\autoref{thm:main}, (iv) implies (ii) because core allocations satisfy
IR and PR, and (ii) implies (iii).
\end{proof}

\section{Uniqueness proofs}\label{app:uniqueness}

At most one rule satisfying SP, IR, and first-choice vacancy-priority
respect executes every top-choice incumbent cycle
(Lemmas~\ref{lem:min-disagreement} and~\ref{lem:unanimityuniqueness}).
On an exchange graph without an even hole, SP, IR, and first-choice
exchange-pair rationality force every such cycle to be executed
(\autoref{lem:topcycleforcing}).
A conflict-free policy with an even hole admits at least two SP, IR, and PR rules
(\autoref{lem:holemultiplicity}).

\begin{lemma}\label{lem:min-disagreement}
Fix any compatibility graph and priority profile, and let
$\varphi$ and $\varphi'$ be distinct rules satisfying SP, IR, and
first-choice vacancy-priority respect. At a profile of disagreement minimizing the number of
objects that applicants rank above their outside options, the two rules
assign every vacancy alike, and one of them executes a compatible
incumbent cycle in which every member receives her first choice, while
the other leaves every member of that cycle with her initial position.
\end{lemma}

The minimum-disagreement argument follows
\citet{Sethuraman2016} and \citet{EkiciSethuraman2024}.

\begin{proof}
Choose a profile $P$ at which $\varphi$ and $\varphi'$ disagree and which minimizes
$s(P)=\sum_{a\in \mathcal{A}}|\{x\in O_a:x\mathrel{P_a} o_a\}|$, the
total number of objects ranked above outside options.

First, each disagreeing applicant $a$ finds exactly one object other
than $o_a$ acceptable, and her two assignments are that object and
$o_a$. Label the rules so that $x=\varphi_a(P)\mathrel{P_a}
\varphi'_a(P)$, so $x\mathrel{P_a} o_a$ by IR. If $a$ found another
object acceptable, let her report $S_a$, which ranks $x$ first, $o_a$
second, and all other objects below $o_a$. Strategy-proofness of
$\varphi$ gives $\varphi_a(S_a,P_{-a})=x$, since otherwise an applicant
with true preferences $S_a$ could report $P_a$ and obtain her first
choice. Strategy-proofness of $\varphi'$ at $P$ rules out
$\varphi'_a(S_a,P_{-a})=x$, so $\varphi'_a(S_a,P_{-a})=o_a$ by IR, and the
rules would disagree at a profile with smaller $s$. IR now gives the
claim.

Second, no disagreeing applicant receives a vacancy under either rule.
Suppose that $\varphi_a(P)=v\in \mathcal{V}$ and $\varphi'_a(P)=o_a$. By
the first step, $a$ ranks $v$ first, so first-choice vacancy-priority
respect for $\varphi'$ requires $\varphi'_b(P)=v$ for some applicant
$b\succ_v a$. Then $\varphi_b(P)\ne v$, so $b$ also disagrees, and by
the first step $b$ ranks $v$ first and $\varphi_b(P)=o_b$. Thus $b$ is denied
her first choice $v$ under $\varphi$, whose recipient $a$ has lower
priority, violating first-choice vacancy-priority respect for $\varphi$.
Interchanging the rules proves the claim. Consequently, the two rules
agree on every vacancy, and every disagreeing applicant is an incumbent,
since a newcomer's only objects are $\varnothing$ and vacancies.

Finally, choose a disagreeing incumbent $i$ with $\varphi_i(P)\ne i$,
interchanging the rules if necessary. By the first two steps,
$\varphi'_i(P)=i$ and $\varphi_i(P)$ is another incumbent's position, so
by \autoref{lem:structure} $i$ belongs to an exchange cycle of incumbents
in $\varphi(P)$. Her predecessor $k$ on that cycle has $\varphi_k(P)=i$, and
injectivity gives $\varphi'_k(P)\ne i$, so $k$ also disagrees. By the
first step, $\varphi'_k(P)=k$ and $\varphi_k(P)$ is her only acceptable
object other than her initial position. Continuing backward around the cycle,
every member receives her first choice under $\varphi$ and keeps her
initial position under $\varphi'$.
\end{proof}

\begin{proof}[Proof of \autoref{rem:capacities}]
DA satisfies these properties. Suppose two such rules differ, and choose
a disagreement profile minimizing the total number of schools ranked
above outside options. The first step of \autoref{lem:min-disagreement}
applies unchanged: every disagreeing student has exactly one acceptable
school and receives it under one rule and her outside option under the
other. Suppose $\varphi_a(P)=h$ and $\psi_a(P)=o_a$. Under $\psi$,
first-choice stability requires every seat at $h$ to go to a student who
outranks $a$. Since $\varphi$ assigns one of those seats to $a$, some
student $b$ assigned to $h$ under $\psi$ must receive a different
assignment under $\varphi$. Student $b$ also disagrees, so $h$ is her
only acceptable school and
$\varphi_b(P)=o_b$. But $\varphi$ gives a seat at $h$ to $a$, below
$b$ in priority, contradicting first-choice stability. Uniqueness
concerns student-to-school assignments, not labels on identical seats.
\end{proof}

\begin{lemma}
\label{lem:unanimityuniqueness}
Fix any compatibility graph and priority profile. At most one rule
satisfying SP, IR, and first-choice vacancy-priority respect executes
every top-choice incumbent cycle at every preference profile.
\end{lemma}

\begin{proof}
If two such rules differed, \autoref{lem:min-disagreement} would give a
profile at which one leaves a top-choice cycle unexecuted, a contradiction.
\end{proof}

\begin{proof}[Proof of \autoref{thm:uniqueness}(iii)]
By \autoref{thm:uniqueness}(i), $\varphi^*$ satisfies SP, IR, and
first-choice vacancy-priority respect and selects from the core.
It executes every top-choice incumbent cycle because an unexecuted
one would block the allocation.
\autoref{lem:unanimityuniqueness} completes the proof.
\end{proof}

A \emph{chord} of a cycle is an edge joining two nonconsecutive vertices
of the cycle.

\AddToHook{para/end}[proofwrap]{\looseness=-1\relax}
\begin{lemma}\label{lem:topcycleforcing}
Suppose the exchange graph $G[\mathcal{I}]$ has no even hole. Every rule
satisfying SP, IR, and first-choice exchange-pair rationality executes every
top-choice incumbent cycle: if $\Gamma=(i_0,\ldots,i_{k-1})$ consists of
distinct incumbents and each $i_t$ ranks $i_{t+1}$ first in $O_{i_t}$, with
subscripts read cyclically, the rule assigns $i_{t+1}$ to $i_t$. This
holds for arbitrary fixed vacancy priorities and arbitrary reports by
applicants outside $\Gamma$, newcomers included.
\end{lemma}

\begin{proof}
Write $\varphi$ for the rule. A two-member top-choice cycle must be
executed: otherwise a denied member and her partner, who weakly prefers
their exchange to her assignment, violate first-choice exchange-pair
rationality. Throughout the proof, every blocking pair has this form.
All reports not explicitly changed, and all vacancy priorities, remain
fixed. If an incumbent is denied her first choice, changing the rest of
her ranking while keeping that first choice unchanged cannot give it to
her. Otherwise, the change would be a profitable misreport at the
original profile.

\noindent
\emph{Moving initial positions to second place.}
Suppose a top-choice cycle is not executed. Starting with a member
denied her successor's position, move her initial position to second place,
preserving all other comparisons. SP prevents her from obtaining her
unchanged first choice, so IR requires her to keep her initial position. Her predecessor
is now denied her first choice. Repeat backward around the cycle.
At the resulting profile, each member ranks only her successor's
position above her initial position. The last reporter retains her
initial position, so injectivity forces every member to do so.

In the reports below, each cycle member ranks all unlisted objects,
including vacancies and positions outside the cycle, below her initial
position. By IR and injectivity, the members receive a permutation of
their own positions whatever outsiders report. No unused edge can
enter these assignments or resolve a blocking pair.

\noindent
\emph{Odd cycles.}
Suppose an odd top-choice cycle $\Gamma=(0,\ldots,k-1)$, $k\ge3$, is
unexecuted. Move each initial position to second place as above. Then, in
order $0,1,\ldots,k-1$, change each
member's preference order to successor's position, predecessor's position,
initial position. After $t$
changes, $1\le t<k$, the only nontrivial IR exchange cycles are
$\Gamma$ and reciprocal pairs on the path $k-1,0,\ldots,t-1$.
The reverse cycle is not yet IR, and chords are unacceptable.
Conditional on excluding $\Gamma$, first-choice protection forces
backward pairing from the terminal vertex $t-1$: its predecessor
claims it first, and the terminal vertex can receive only her own
position or that predecessor's. Repeat on the remaining path.

Inductively, SP excludes $\Gamma$ at every change. For $t<k$, the
reporter was isolated in the previous pair graph and retained her
initial position. Before the last change, backward pairing on the path of
all $k$ members leaves member $k-1$ unmatched because $k$ is odd. Thus
every reporter previously kept her initial position and retains the same first
choice, which SP prevents her from obtaining. At the final profile,
the reverse cycle also fails first-choice protection: it denies $i$
position $i+1$ while giving $i+1$ position $i$, so this pair blocks
with one equality. A pair allocation leaves some member unmatched,
and she strictly prefers exchanging with her predecessor, who claims
her position first. No allocation remains. Hence every odd top-choice
cycle is executed.

\noindent
\emph{An even cycle with an odd-splitting chord.}
Suppose an unexecuted even top-choice cycle $\Gamma$ has a chord with
endpoints at even cyclic distance. Move each initial position to second place
as above and label the forward paths between the chord's endpoints
$a,u_1,\ldots,u_L,b$ and $b,z_1,\ldots,z_M,a$, where $L,M$ are
positive odd integers. Let $Q=(a,b,z_1,\ldots,z_M)$ be the odd
shortcut cycle. Only edges of $\Gamma$ and this chord will be
acceptable.

First let $b$ rank $z_1,u_L,b$. Since she previously retained her
initial position, SP excludes $\Gamma$. The only remaining nontrivial IR
exchange is $\{u_L,b\}$, forced because $u_L$ claims $b$ first and
$b$ can otherwise only retain her initial position. All other members retain
theirs. Now let $a$ report $u_1$ first, $b$ second, $a$ third,
and all other objects below $a$.
At this report, SP again excludes $\Gamma$. Yet $a$
must receive $u_1$ or $b$:
reporting $b$ first creates the odd top-choice cycle $Q$, which must
be executed. Since $u_1$ accepts only her successor, assigning $u_1$
to $a$ would require $\Gamma$. Thus $Q$ is executed and every $u_j$
retains her initial position.

For $t=1,\ldots,L$, let $u_t$ switch to successor's position, predecessor's position,
initial position, leaving $a,b$, and the $z_j$ unchanged. Denote the resulting
profiles by $P^{(t)}$, with $P^{(0)}$ the profile just constructed.
At every stage, $a$ can still obtain $b$ by reporting it first, so
SP requires her assignment to be $u_1$ or $b$.

We first classify allocations conditional on excluding $\Gamma$. This
classification will show that the next reporter retains her initial position,
so SP continues to exclude $\Gamma$ after her report changes. Besides
$\Gamma$ and $Q$, the forward chord cycle
$(a,u_1,\ldots,u_L,b)$ gives $b$ the unacceptable position $a$,
and the reverses of $\Gamma$ and $Q$ give $a$ the unacceptable
position $z_M$. The reverse chord cycle becomes IR only at $t=L$,
but then $a$ is denied $u_1$ while $u_1$ receives $a$, violating
first-choice protection. Thus only $Q$, reciprocal pairs, and
retentions remain.

If $Q$ is executed at $P^{(t)}$, the remaining pair graph is the path
$u_1,\ldots,u_t$ and isolated later members. Backward pairing is
forced as above. For odd $t$, it leaves $u_1$ unmatched, so
$\{a,u_1\}$ blocks. For even $t$, it pairs the whole path and leaves
every later $u_j$ with her initial position.

If only pairs are used and $t<L$, the pair graph consists of
$a,u_1,\ldots,u_t$, the separate pair $\{u_L,b\}$, and isolated
vertices. Backward pairing is forced. Even $t$ leaves $a$ unmatched,
contrary to her attainable option $b$. Odd $t$ pairs the whole first
path, giving $a$ position $u_1$, and leaves later $u_j$, other than
$u_L$, with their initial positions. At $t=L$, the pair graph is
$a,u_1,\ldots,u_L,b$. Its odd number $L+2$ of vertices forces $a$
to remain unmatched when pairing backward from $b$. This is also
impossible, and $Q$ is excluded because $L$ is odd.

Finally, $\Gamma$ is excluded at every stage. It is excluded at
$P^{(0)}$. Inductively, at $P^{(t-1)}$ the preceding classification leaves
$u_t$ with her initial position. If $t$ is odd, then $t-1$ is even, so
the $Q$ pattern applies and leaves $u_t$ with her initial position. If
$t$ is even, the pair pattern applies. Since $L$ is odd, $t<L$, so $u_t$
lies neither on the first path $a,u_1,\ldots,u_{t-1}$ nor in
$\{u_L,b\}$ and again retains her initial position. Her first choice is unchanged, so SP again rules out
$\Gamma$. The classification therefore applies throughout and leaves
no allocation at $t=L$, a contradiction.

\noindent
\emph{Even-hole-freeness.}
Every even cycle of length at least four in an even-hole-free graph
has a chord joining vertices at even cyclic distance. Otherwise its
induced subgraph would be bipartite, with the cycle's two parity
classes as parts. A shortest cycle in that subgraph would be chordless
and even, hence an even hole in $G[\mathcal I]$. Together with the
two-member and odd-cycle arguments, this covers every top-choice
incumbent cycle.
\end{proof}

\begin{lemma}\label{lem:holemultiplicity}
Suppose $(G,\succ)$ is conflict-free. If $G[\mathcal{I}]$ has an even
hole, there are at least two SP, IR, and PR rules. In particular, these three properties do not
single out PTR.
\end{lemma}
\begin{proof}
Let $H$ induce an even hole in $G[\mathcal I]$. We first construct a
rule on $H$, add its eligible vacancies, and extend it to the remaining
applicants, whom we call \emph{outsiders}.

\emph{A second rule on the hole.}
Delete each edge unless both endpoints prefer each other's position to
their own, and run TTC on the remaining exchange graph. Call this rule $\psi$.
Except at the two profiles where all members find both neighbors
acceptable and rank their first choices consistently around the hole,
the pair-matching market has a unique stable matching, which $\psi$
selects. If an edge is deleted, the remaining exchange graph consists of paths.
Isolated members retain their initial positions. Among remaining vertices,
every nontrivial path has a mutual-first pair, which must exchange.
Successively removing these forced assignments uniquely determines
the stable matching. If every edge remains, a
mutual-first pair again leaves paths. Otherwise all first choices point
consistently around the hole, giving one of the two exceptions, and
$\psi$ executes that top-choice cycle. Deleted edges cannot block an
IR allocation, so $\psi$ satisfies IR and PR.

For SP, use the hole's bipartition. At nonexceptional profiles the
stable matching is unique, so DA produces $\psi$ whichever parity
class proposes. For a given deviation, take the deviator's class as the
proposing side. DA's strategy-proofness then rules out a gain between
nonexceptional profiles. At an exceptional true profile everyone
already receives her first choice. For a nonexceptional true profile
and an exceptional report by $i$, call her reported first choice her
successor and the other neighbor her predecessor. If her initial position is
truly first, IR already gives it to her. If her predecessor is truly
first, they are mutual first choices and already exchange. The only
remaining nonexceptional order is successor's position, initial position,
predecessor's position.
The mutually acceptable edges then form a path with an even number of
vertices starting at $i$.
Pairing backward from the other endpoint is forced and gives $i$ her
successor. The exceptional report cannot improve her assignment.

\emph{Adding vacancies.}
The policy is conflict-free, so it admits an SP, IR, and PR rule by
\autoref{thm:main}. All members of $H$ belong to one exchange component.
The necessity direction of \autoref{prop:policy} therefore implies that
each vacancy has at most one eligible member of $H$. Thus the sets
$\mathcal V_i=N_G(i)\cap\mathcal V$, $i\in H$, are disjoint.
Let $y_i$ be $i$'s favorite object in $\{i\}\cup\mathcal V_i$.
Define a local preference $\tilde P_i$ over her own label and her two
neighbors by placing her own label where she ranks $y_i$, preserving
the neighbors' order. This preference depends only on $P_i$.
Run $\rho=\psi(\tilde P)$, give $i$ position $\rho_i$ if
$\rho_i\ne i$, and give her $y_i$ otherwise, withdrawing her
initial position when $y_i$ is a vacancy. Denote the resulting assignments
by $\psi^+$.

These assignments are feasible: $\rho$ permutes the positions in $H$,
the vacancy sets are disjoint, and a member receiving a vacancy has her
own label under $\rho$, so nobody else receives her withdrawn position.
Local IR gives each member an object weakly preferred to $y_i$, hence
to her initial position and every eligible vacancy. Consequently no member
has a vacancy claim. Any blocking exchange pair within $H$ would also
block $\rho$, since replacing the own label by $y_i$ preserves the
relevant comparisons. There are no chords in $H$.

To verify SP despite the report-dependent $y_i$, fix the other reports.
A deviation by $i$ leaves the other members' local preferences fixed,
but it can change both $\tilde P_i$ and $y_i$. If it gives $i$ a neighbor's
position, local SP says that her truthful assignment is weakly preferred
to that position. If it gives her own label locally, her actual
assignment is some $y'_i\in\{i\}\cup\mathcal V_i$. By local IR and
the definition of $y_i$, her truthful assignment is weakly preferred to
$y_i$, which is weakly preferred to $y'_i$. Thus $\psi^+$ is SP for
members of $H$.

\emph{Extension to the full market.}
Call a profile \emph{quiet} if every outsider ranks her outside option
first. At a quiet profile, let $\varphi'$ give outsiders their outside
options and members of $H$ their $\psi^+$ assignments, leaving all
other vacancies unassigned. At other profiles let
$\varphi'=\varphi^*$.

At quiet profiles feasibility and IR follow by construction.
Members of $H$ have no vacancy claim or blocking pair within $H$, and
outsiders receive their first choices, so they neither claim vacancies
nor participate in blocking exchanges. At other profiles PTR gives IR
and PR.

A member of $H$ cannot switch branches by misreporting, and both
branches are SP for her. Each outsider receives her PTR assignment at
every profile, since IR gives her her first-choice outside option at
quiet profiles under PTR as well. Thus even a report that switches
branches cannot benefit her, and $\varphi'$ is SP.

Finally, let every outsider rank her outside option first and every
member of $H$ rank her successor first, her initial position second, her
predecessor third, and all other objects below her initial position. No hole
edge is mutually acceptable, so $\varphi'$ leaves every member in place.
PTR instead executes their top-choice cycle. The rules differ.
\end{proof}

\RemoveFromHook{para/end}[proofwrap]
\begin{proof}[Proof of \autoref{thm:uniqueness}(ii)]
By \autoref{thm:uniqueness}(i), $\varphi^*$ satisfies SP, IR, and PR,
hence first-choice PR, and executes every top-choice cycle.
If $G[\mathcal{I}]$ has no even hole, \autoref{lem:topcycleforcing}
shows that every SP, IR, and first-choice PR rule executes every
top-choice cycle, and \autoref{lem:unanimityuniqueness} gives uniqueness.
If $G[\mathcal{I}]$ has an even hole, \autoref{lem:holemultiplicity}
supplies a second SP, IR, and PR rule for the same priorities. Both
rules satisfy first-choice PR.
\end{proof}

\ifsupplement
\clearpage
\renewcommand{\thesection}{S}
\renewcommand{\sectionname}{}
\numberwithin{equation}{section}
\numberwithin{figure}{section}
\numberwithin{table}{section}
\section{Supplemental Appendix}\label{app:supplementary}

We use the notation of \bodysecref{sec:model} and \appref{app:slots}.

\subsection{Guarantees under arbitrary policies}\label{app:robust-proofs}

Recall that $\mu^{-1}(v)$ denotes the recipient of an assigned vacancy
$v$ under allocation $\mu$. At Step~3 of the algorithm, $r(a)$ denotes
the vacancy at the end of active applicant $a$'s pointer path. The sink
components are the finalization groups of Step~3.

\begin{proof}[Proof of \autoref{prop:robust}]
Fix a policy and a preference profile, and let $\mu$ be the allocation
produced by the algorithm.
The termination, feasibility, IR, incumbent-cycle, and non-wastefulness
arguments in the proof of \autoref{lem:existence} do not require the
policy to be conflict-free. They also establish that an applicant held
by a vacancy keeps pointing to it until rejection or finalization, that
each new tentative recipient has higher priority than the previous one,
and that the final recipient
of a vacancy outranks everyone it rejected. An applicant's pointer moves
only to less-preferred objects, because her options only shrink.

\emph{Part (i).}
An applicant who ranks a vacancy first points to it at the start of
Step~1. If she is ever rejected, its eventual recipient outranks her. If
she is never rejected, she keeps pointing to it until finalization and
receives it. This proves first-choice vacancy-priority respect, and the absence
of a blocking compatible incumbent cycle gives first-choice
exchange-pair rationality.

\noindent
\emph{Part (ii).}
Suppose $a$ has justified envy of $v$, and let $b=\mu^{-1}(v)$.
Vacancy $v$ never rejected $a$, since its final recipient would then
outrank her. Nor can $a$ depart before or together with $v$, because the
active, unrejected vacancy would be available when she chose her
assignment. Let $S$ be the sink component containing $v$ at its
finalization. Applicant $a$ is still active. Since $a\succ_v b$, the
dependency graph contains the link $v\to r(a)$, so $r(a)\in S$.
If $a$ pointed to a
vacancy, it would be $r(a)$, and she would be held by it and depart in
the same finalization. Since no newcomer points to $\varnothing$ and no
incumbent to her own position at Step~3, $a$ therefore points to
another incumbent's position $j$. In particular $a$ is an incumbent,
because a newcomer points only to $\varnothing$ or a vacancy, and the
availability of $v$ gives $j\mathrel{P_a}v$.

Use the expanded dependency graph from the proof of \autoref{lem:safe}: applicants
point to their current choices, and each vacancy holding an applicant points to the
active applicants it ranks above its tentative recipient. There is an arc $v\to a$
and a pointer path from $a$ to $r(a)$, and a dependency path within $S$
from $r(a)$ to $v$ lifts to a return path in the expanded dependency graph. Erasing
loops gives a directed path from $a$ to $v$, which closes with
$v\to a$. Its first arc is $a\to j$, the only outgoing arc of $a$,
and its final arc is $b\to v$, because $b$ is the sole applicant
pointing to $v$. Every vacancy on the
cycle is entered from the applicant it holds and exits toward a higher-priority
applicant, and each arc joins vertices adjacent in $G$, so the cycle is
priority-consistent, and the incumbent--incumbent arc $a\to j$ makes it
a conflict cycle, including when $j=b$. Every vacancy on this cycle
belongs to $S$ and is assigned to its predecessor at this finalization.
Each applicant on the cycle points to her successor at that step and
weakly prefers that object to her final assignment because her
available options can only shrink.
\end{proof}

\begin{proof}[Proof of \autoref{thm:robust-exact-priorities}]
Necessity follows from \autoref{prop:robust}(ii). For sufficiency,
orient a cycle as in (ii), and let $j$ be the successor of $a$, allowing
$j=b$. Since $j$ is an incumbent and the edge $\{a,j\}$ joins two
applicants, $a$ is an incumbent. Every other applicant on the cycle
ranks her successor first and her outside option second, with all
remaining objects below. Incumbent $a$ ranks $j\mathrel{P_a}v\mathrel{P_a}a$
and puts all remaining objects below her initial position. Every applicant
outside the cycle ranks her outside option first. Thus $a$ ranks exactly
two objects above her outside option, and every other applicant ranks at
most one.

On-cycle applicants point to their successors, and outside applicants
leave in Step~2 before any finalization, newcomers unmatched and
incumbents with their positions. Each on-cycle vacancy holds its
predecessor, no applicant points to any other vacancy, no on-cycle newcomer
points to $\varnothing$, and the on-cycle incumbents' pointers form no
cycle, because the only cycle they follow contains a vacancy. For each
on-cycle vacancy $w$, priority consistency makes its successor outrank
its tentative recipient, and that successor's pointer path ends at the next vacancy
along the cycle. The dependency graph therefore contains a directed
cycle through all its vertices, the on-cycle vacancies, which form one
sink strongly connected component, since every active applicant lies on
the cycle and further eligibility edges only add links within it.

All these vacancies are finalized together, each to its cycle
predecessor. In particular, $\mu_b=v$. The surviving applicants are
incumbents, since every on-cycle newcomer is held by the vacancy
she points to. They form forward paths whose last successor positions
have been removed, and apart from their own positions they find
acceptable only their forward successors' positions. Incumbent $a$'s
second choice $v$ has also been removed. These paths unwind, each
incumbent in turn pointing to her own position, so every survivor,
including $a$, retains her initial position. Thus $v\mathrel{P_a}\mu_a=a$, and
every other applicant receives her sole acceptable vacancy or has no
vacancy above her outside option, so $(a,v)$ is the only justified-envy pair.

Interchange $a$'s first two choices, leaving all other comparisons
unchanged. She immediately points to $v$, which ranks her above $b$.
No applicant other than $a$ and $b$ finds $v$ acceptable, and since
outside options remain available until departure, no other applicant
can ever point to $v$. Incumbent $a$ is never rejected and eventually
receives $v$, improving on her true assignment $a$.
\end{proof}

\begin{proof}[Proof of \autoref{cor:robust-sp}]
Under a conflict-free policy, \autoref{thm:uniqueness}(i) gives SP and full
vacancy-priority respect. Conversely, orient any conflict cycle
consistently with priorities. Because it has an exchange
edge and a vacancy, some vacancy $v$ is followed by at least two
applicants. Let $i$ be the first of them and $b$ the predecessor of $v$.
The successor of $i$ is an applicant joined to $i$ by an edge, hence an
incumbent. At the profile constructed in the proof of
\autoref{thm:robust-exact-priorities}, PTR violates $i$'s priority at
$v$, and $i$ gains by interchanging her first two choices.
\end{proof}

\emph{Carol's deviation in \bodysecref{sec:robustness}.}
Start from \autoref{fig:policy-cycle} and make Ann eligible for $w$ as
well. Ann and Bob are incumbents, and Carol is a newcomer. The vacancy
priorities are $\text{Ann}\succ_v\text{Carol}$ and
$\text{Bob}\succ_w\text{Carol}\succ_w\text{Ann}$. Let
\[
P_A: w,v,\text{Ann},\text{Bob}
\qquad
P_B: \text{Ann},w,\text{Bob}
\qquad
P_C: w,v,\varnothing.
\]

Under truthful reporting, $w$ holds Carol and rejects Ann, who points
to $v$. Bob points to Ann and outranks Carol at $w$, giving the link
$w\to v$. Finalizing $v$ first withdraws Ann's position, so Bob displaces
Carol at $w$. Ann receives $v$, Bob receives $w$, and Carol remains
unmatched. No applicant has justified envy.

If Carol reports $v$ first, $v$ holds Carol and $w$ holds Ann. Ann
outranks Carol at $v$, and Carol outranks Ann at $w$, so Step~3
finalizes both vacancies together. Carol receives $v$, which she
prefers to remaining unmatched, and Bob keeps his position. Under the
reports, only Bob has justified envy at $w$. Under their true
preferences, both Bob and Carol have justified envy at $w$.

\begin{proof}[Proof of \autoref{prop:ptr-undominated}]
\emph{Part (i).} Fix a profile $P$ and write $\mu=\varphi^*(P;\succ)$.
For an allocation $\mu$ and a set $S\subseteq\mathcal{A}$ of applicants,
$\mu(S)=\{\mu_a:a\in S\}$ denotes the set of objects they receive.
For an allocation $\lambda$, let $\mathcal J(P,\lambda;\succ)$ denote
the set of justified-envy pairs at assigned vacancies.
Let $\nu$ Pareto dominate $\mu$, and put
$\Delta=\{a:\nu_a\mathrel{P_a}\mu_a\}\ne\varnothing$. By strict
preferences, every applicant outside $\Delta$ receives the same object
under both allocations. Every $a\in\Delta$ receives a vacancy under
$\nu$: her outside option cannot strictly improve an IR assignment,
which settles the case of a newcomer, and if an incumbent $a\in\Delta$
received another incumbent's position, \autoref{lem:structure} would
place her in an exchange cycle of incumbents under $\nu$ that weakly
improves all its members relative to $\mu$ and strictly improves $a$,
contradicting \autoref{prop:robust}(i).

Non-wastefulness of $\mu$ implies that $\nu_a$ is assigned under $\mu$ for
every $a\in\Delta$. Define $f(a)=\mu^{-1}(\nu_a)$. Since $a$ changes
assignments, $f(a)\ne a$, and injectivity of $\nu$ on vacancies gives
$\nu_{f(a)}\ne\nu_a=\mu_{f(a)}$, so strict preferences and Pareto
dominance imply $f(a)\in\Delta$. The map $f:\Delta\to\Delta$ is
injective, hence bijective, so $\mu(\Delta)=\nu(\Delta)\subseteq\mathcal
V$: every member of $\Delta$ receives a vacancy under both allocations.
In particular, a newcomer who is unmatched under $\mu$ is not in
$\Delta$ and remains unmatched under $\nu$.

Consider the first event at which an applicant $a\in\Delta$ is rejected
by her target vacancy $\nu_a$ or a member of $\Delta$ departs. Such an
event exists because every applicant departs. A departure cannot be
first: a departing $a$ receives $\mu_a$, which she ranks below $\nu_a$,
so her target had already rejected her or departed with its final
recipient $f(a)\in\Delta$. Objects finalized together remain available
when their recipients choose their pointers, so simultaneous
finalizations do not change this conclusion. The first event is
therefore the rejection of some $a\in\Delta$ by $v=\nu_a$. Let $b$ be
the applicant held by $v$ at this rejection, so $b\succ_v a$, and
$\nu_b\ne v$ by injectivity. If $\nu_b\mathrel{P_b}v$, pointer
monotonicity gives $\nu_b\mathrel{P_b}v\mathrel{R_b}\mu_b$, so
$b\in\Delta$, $\nu_b$ is a vacancy, and when $b$ chose $v$ the better
vacancy $\nu_b$ had already rejected her or departed with
$f(b)\in\Delta$. Either case gives an earlier rejection by a target
vacancy or departure from $\Delta$, a contradiction. Rejections in a batch can
be ordered arbitrarily, since the applicants involved chose their
pointers before the batch. Thus $v\mathrel{P_b}\nu_b$, and $(b,v)\in\mathcal J(P,\nu;\succ)$.
The final recipient of $v$ under $\mu$ is $b$ or an applicant outranking
$b$, because each new tentative recipient has higher priority than the
previous one, so
$(b,v)\notin\mathcal J(P,\mu;\succ)$, proving the strict set difference.
Since applicants outside $\Delta$ retain their assignments, the equality
$\mu(\Delta)=\nu(\Delta)\subseteq\mathcal V$ proves the stated
invariance properties.

\emph{Part (ii).} Suppose $\psi$ Pareto dominates PTR. Choose $P$ and
$a$ for whom the improvement is strict. By part~(i),
$x=\psi_a(P;\succ)$ and $\varphi_a^*(P;\succ)$ are distinct vacancies,
with $x\mathrel{P_a}\varphi_a^*(P;\succ)\mathrel{P_a}o_a$.
Raise $o_a$ to immediately below $x$,
preserving the order of all other objects, and call the resulting
profile $Q$. Compare the runs of PTR at $P$ and $Q$. If $a$ always
pointed to the same object in both runs, she would receive the same
vacancy at $Q$ as at $P$, although she now ranks it below $o_a$.
IR rules this out. When her pointers first differ, the available
objects are the same in both runs. Since she has raised only $o_a$,
she points to $o_a$ at $Q$ and receives it in Step~2. Since $\psi$
weakly improves every applicant's assignment at $Q$, part~(i) and
strict preferences imply $\psi_a(Q;\succ)=o_a$ as well. At the true profile $Q$, reporting
$P_a$ gives applicant $a$ the vacancy $x$, which she prefers to $o_a$.
This profitable report lowers $o_a$ and preserves the relative order of
all other objects.
\end{proof}

\subsection{Constrained efficiency}\label{app:compatibility-uniqueness}

\begin{proof}[Proof of \autoref{cor:constrained}]
PTR is SP by \autoref{thm:uniqueness}(i). Let $\mu$ be its applicant-optimal core
allocation. If an IR and PR allocation $\nu$ Pareto dominated $\mu$, any
compatible incumbent cycle blocking $\nu$ would also block $\mu$. Hence
$\nu$ would belong to the core, contradicting applicant optimality.
Thus PTR is constrained efficient. Conversely, let $\psi$ be SP and constrained
efficient with $\psi\ne\varphi^*$. Both rules satisfy SP, IR, and
first-choice PR, so by \autoref{lem:min-disagreement} there is a profile
$P$ at which one of them executes a compatible incumbent cycle $\Gamma$
giving every member her first choice, while the other, $\rho$, leaves
every member of $\Gamma$ with her initial position. Let
$\mu$ execute $\Gamma$ and agree with $\rho(P)$ on every other
applicant. Since $\rho(P)$ gives each member of $\Gamma$ her own
position, nobody else receives these positions under $\rho(P)$ and no
member receives a vacancy, so $\mu$ is an allocation by
\autoref{lem:structure}, and it is IR. A member $c$ of $\Gamma$ weakly
prefers another incumbent's position to $\mu_c$ only if that position
is $\mu_c$, whose owner also receives her first choice, so no pair
involving $\Gamma$ blocks $\mu$, and members of $\Gamma$ have no vacancy
claims. Every other applicant's assignment and every vacancy's recipient
are as in $\rho(P)$, which is PR. Hence $\mu$ is IR and PR and Pareto
dominates $\rho(P)$, since members of $\Gamma$
strictly gain and nobody else changes, contradicting the constrained
efficiency of $\rho$, which is $\psi$ or $\varphi^*$.
\end{proof}

\subsection{Exchange components, priority reform, and filling vacancies}\label{app:priority-free}

\begin{proof}[Proof of \autoref{prop:componentcompletion}]
Complete each exchange component of $G$ by permitting every pair in it to
exchange, and denote the resulting graph by $G^+$. The same operation on
$G'$ gives the same graph, so it suffices to prove that $G$ and $G^+$ are
conflict-free for the same priority profiles.
Completion adds only exchange edges, since every exchange component
with two or more members consists of incumbents. Every conflict cycle
in $G$ therefore remains one in $G^+$.

Conversely, choose a conflict cycle in $G^+$ minimizing the number of
added edges it uses. Suppose it uses an added edge $\{i',j'\}$ inside
an exchange component $K$ of $G$. Since the cycle contains a vacancy,
it must leave $K$ and return. Follow it from an incumbent $i\in K$
where it leaves to the next incumbent $j\in K$ where it returns.
This gives an oriented path between distinct incumbents whose internal
vertices lie outside $K$.

Close this path with a path from $j$ to $i$ in $G[K]$, using
only original exchange edges. The paths have disjoint interiors, so
their union is a cycle containing a vacancy and an exchange edge. Its vacancy
comparisons are unchanged, making it a conflict cycle. It omits the
added edge $\{i',j'\}$ and introduces no new added edges, contradicting
the choice of the original cycle.
\end{proof}

\emph{Common seniority across vacancies.}
In \autoref{fig:policy-cycle}, Ann and Bob may exchange, and Carol is a
newcomer. Vacancy $v$ is open to Ann and Carol, and $w$ to Bob and Carol.
Each vacancy thus has one eligible applicant from each component.

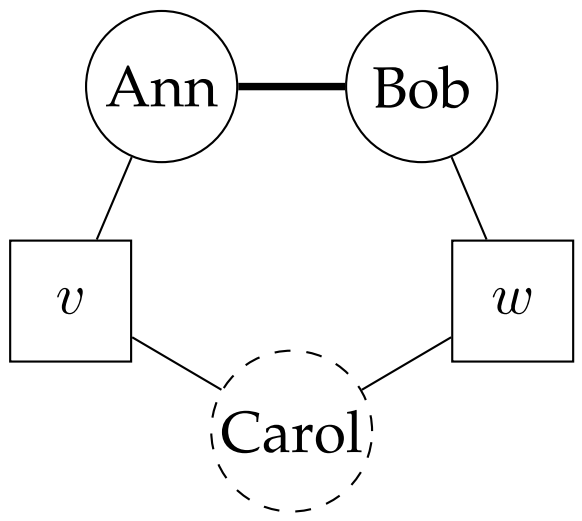
\begin{figure}[H]
\centering
\begin{minipage}[c]{.34\linewidth}
\centering
\begin{tikzpicture}[scale=.86, every node/.style={font=\small}]
\node[draw,circle,minimum size=10mm,inner sep=1pt] (ann) at (-1,1) {Ann};
\node[draw,circle,minimum size=10mm,inner sep=1pt] (bob) at (1,1) {Bob};
\node[draw,rectangle,minimum size=8mm] (w) at (1.7,-.65) {$w$};
\node[draw,circle,dashed,minimum size=10mm,inner sep=1pt] (carol) at (0,-1.65) {Carol};
\node[draw,rectangle,minimum size=8mm] (v) at (-1.7,-.65) {$v$};
\draw[line width=1.4pt] (ann)--(bob);
\draw (bob)--(w)--(carol)--(v)--(ann);
\end{tikzpicture}
\end{minipage}\hfill
\begin{minipage}[c]{.64\linewidth}
\centering
\small
\begin{tabular}{@{}llll@{}}
\toprule
Priorities & Priority at $v$ & Priority at $w$ & Implementable\\
\midrule
$\succ^{-}$ & Ann $\succ$ Carol & Bob $\succ$ Carol & Yes\\
$\succ^{\mathrm{conflict}}$ & Ann $\succ$ Carol & Carol $\succ$ Bob & No\\
$\succ^{+}$ & Carol $\succ$ Ann & Carol $\succ$ Bob & Yes\\
\bottomrule
\end{tabular}
\end{minipage}
\caption{Three priority profiles on the same graph. Solid circles denote
incumbents, the dashed circle a newcomer, and squares vacancies. The bold
edge permits exchange. Thin edges indicate eligibility.}
\label{fig:policy-cycle}
\end{figure}

Suppose both vacancies use the same seniority ranking, with Ann first, Carol
second, and Bob third. Then $v$ ranks Ann above Carol, while $w$ ranks Carol
above Bob. Under this profile, $\succ^{\mathrm{conflict}}$, the cycle
$\text{Ann}\to\text{Bob}\to w\to\text{Carol}\to v\to\text{Ann}$ is a
conflict cycle. Reversing the comparison at $w$ gives $\succ^{-}$, and
reversing it at $v$ gives $\succ^{+}$. Either change puts the components
in the same order at both vacancies and makes the policy conflict-free.

\emph{Filling vacancies without priority comparisons.}
Suppose an eligible applicant's first-choice vacancy must be filled but
may go to any eligible applicant. The obstruction in \autoref{prop:policy}
persists even under this weaker requirement.

\begin{proposition}[Exchange rights and filling vacancies]
\label{prop:priority-free}
Fix a compatibility graph $G$. There exists an SP and IR rule that
executes every exchange between compatible incumbents who rank each
other's positions first and fills every vacancy ranked first by at
least one eligible applicant if and only if each vacancy has at most
one eligible applicant in each exchange component.
\end{proposition}

\autoref{tab:priority-free} illustrates necessity in the
simplest case. Ann and Bob may exchange, so they belong to the same
component, and both are eligible for $v$.

\begin{table}[H]
\centering
\caption{Admissible allocations under withdrawal, listed in Ann--Bob
order. At each profile we impose IR, execute a swap if Ann and Bob
rank each other first, and fill $v$ if either ranks it first.}
\label{tab:priority-free}
\begin{tabular}{@{}llll@{}}
\toprule
Profile & Ann's preferences & Bob's preferences & Admissible allocations\\
\midrule
$Q$ & Bob, $v$, Ann & Ann, $v$, Bob & $(\text{Bob},\text{Ann})$\\
$P^0$ & Bob, $v$, Ann & $v$, Ann, Bob & $(\text{Ann},v)$ or $(v,\text{Bob})$\\
$P^1$ & $v$, Bob, Ann & $v$, Ann, Bob & $(\text{Ann},v)$ or $(v,\text{Bob})$\\
$P^2$ & $v$, Bob, Ann & Ann, $v$, Bob & $(\text{Ann},v)$ or $(v,\text{Bob})$\\
\bottomrule
\end{tabular}
\end{table}

At $Q$, Ann and Bob must exchange because each ranks the other's
position first. At $P^0$, Bob instead ranks $v$ first, so it must be
filled. If Ann received $v$, withdrawal would leave Bob in his own
position. He could then obtain Ann's position, which he prefers, by
reporting as at $Q$. Thus SP requires Bob to receive $v$ at $P^0$.
At $P^2$, only Ann's ranking differs from $Q$, so the same argument
requires her to receive $v$.

At $P^1$, both rank $v$ first. If Ann receives $v$ there, she can
obtain it at $P^0$ by reporting as at $P^1$. If Bob receives $v$
there, he can obtain it at $P^2$ by reporting as at $P^1$. In either
case the applicant gains by receiving $v$ instead of her own position,
contradicting SP.
The proof extends the argument along a shortest exchange path between two
incumbents eligible for the same vacancy.

\begin{proof}[Proof of \autoref{prop:priority-free}]
For sufficiency, rank the exchange components in the same order at every
vacancy, as in the proof of \autoref{prop:policy}. Under these priorities
PTR satisfies SP, IR, and PR. Exchange-pair rationality requires the swap
when compatible incumbents rank each other's positions first, and
vacancy-priority respect requires every vacancy ranked first by an eligible
applicant to be filled.

For necessity, suppose an SP and IR rule lets compatible incumbents
exchange whenever they rank each other's positions first and fills every
vacancy ranked first by at least one eligible applicant. Whenever an incumbent $i$ ranks a
compatible incumbent $j$'s position first, $j$ must weakly prefer her
assignment to $i$'s position. Otherwise $j$ could obtain $i$'s
position by reporting it first, since the rule would then execute their
swap. In particular, $j$ cannot retain her own position if she prefers
$i$'s to it.

Suppose a vacancy $v$ has two eligible
applicants in one exchange component. They are incumbents, since every
newcomer is a singleton component. Among the incumbents eligible for
$v$ in that component, choose two at minimum exchange distance and a
shortest path $c_1,\ldots,c_k$ between them, where $k\ge2$. This path
is induced, and no interior incumbent is eligible for $v$.
Put $c_0=c_{k+1}=v$.

Every applicant off the path ranks her outside option first. On the
path, only the two neighbors in the cycle
$(v,c_1,\ldots,c_k)$ are ever ranked above the reporting incumbent's initial position, and all
unlisted objects are ranked below it in a fixed order. IR and
the withdrawal condition then imply that an allocation consists of reciprocal swaps
along the path, retained initial positions, and possibly the assignment of $v$
to an endpoint. No other vacancy is assigned. We call a path incumbent
\emph{unmatched} when she retains her initial position.

For $m=0,\ldots,k$, define $P^m$ by
\[
\begin{array}{ll}
c_{t-1}\mathrel{P^m_{c_t}}c_{t+1}\mathrel{P^m_{c_t}}c_t,
&t\le m,\\[2pt]
c_{t+1}\mathrel{P^m_{c_t}}c_{t-1}\mathrel{P^m_{c_t}}c_t,
&t>m.
\end{array}
\]
At least one endpoint ranks $v$ first, so the rule must assign it.
Call an allocation \emph{type $X$} if $c_1$ receives $v$
and \emph{type $Y$} if $c_k$ receives it.

\medskip\noindent
\emph{The path restricts unmatched incumbents.}
At $P^m$, every unmatched incumbent on the path belongs to
$U_m=\{c_m,c_{m+1}\}\cap\{c_1,\ldots,c_k\}$.
An unmatched $c_t$ with $t\le m-1$ is claimed first by $c_{t+1}$,
and one with $t\ge m+2$ by $c_{t-1}$. In either case $c_t$ prefers
the claimant's position to retaining her own, contrary to the SP
implication above.

Fixing the vacancy recipient and unmatched incumbents uniquely fixes
the consecutive pairs on the remaining path. Each segment before,
between, or after unmatched incumbents must contain an even number of
vertices, because its members exchange in pairs. Under type $X$, this path
has indices $2,\ldots,k$, so a lone unmatched index must be even, and
two unmatched indices must be even followed by odd. Under type $Y$,
the path is $1,\ldots,k-1$, so the corresponding parities are odd,
and odd followed by even.

\medskip\noindent
\emph{An even number of incumbents.}
Suppose $k$ is even. After assigning $v$, an odd number of incumbents
remain, so exactly one member of $U_m$ is unmatched. A type-$X$
allocation leaves the even-indexed member of $U_m$ unmatched, and a
type-$Y$ allocation leaves its odd-indexed member unmatched. Thus
$P^0$ admits only type $Y$, while $P^k$ admits only type $X$.

Only $c_m$ changes her report between $P^{m-1}$ and $P^m$. If $m$ is
odd, she retains her initial position under type $Y$ at both profiles, and
receives a neighbor under type $X$. If $m$ is even, the roles of the
types are reversed. Both neighbors are above her initial position in both
reports. Therefore a change of type would let $c_m$ profitably
misreport at the profile where she retains her initial position. SP preserves
the type along the entire sequence, contradicting its different types
at $P^0$ and $P^k$.

\medskip\noindent
\emph{An odd number of incumbents.}
Suppose $k$ is odd. Let $X^*$ and $Y^*$ denote the two allocations
with no unmatched incumbent:
\[
\begin{array}{ll}
X^*:&c_1\gets v,\quad (c_2,c_3),(c_4,c_5),\ldots,(c_{k-1},c_k),\\
Y^*:&(c_1,c_2),(c_3,c_4),\ldots,(c_{k-2},c_{k-1}),\quad c_k\gets v.
\end{array}
\]
Here each parenthesized pair exchanges initial positions. Since $k-1$ incumbents remain after
assigning $v$, any other admissible allocation at $P^m$ leaves exactly
$c_m,c_{m+1}$ unmatched. The parity calculation shows that this
\emph{deficient} allocation can be type $X$ only for even $m$, and
type $Y$ only for odd $m$. It is unavailable at $m=0$ or $m=k$.

First the rule must choose $Y^*$ at $P^0$. Let $c_1$ instead report
$c_2$ first, $c_1$ second, and $v$ below her initial position. IR excludes
assigning $v$ to $c_1$, while $c_k$ still ranks $v$ first. The only
possible unmatched incumbent remains $c_1$, so parity excludes any
unmatched incumbent and forces $Y^*$. Because $Y^*$ gives $c_1$ her first
choice at $P^0$, SP forces that assignment at $P^0$ as well, and hence
forces $Y^*$. By reflection, refusing $v$ at $c_k$ forces $X^*$ at $P^k$.

We now propagate $Y^*$ from $P^0$ to $P^{k-1}$.
Suppose the rule chooses $Y^*$ at $P^{m-1}$, where $1\le m\le k-1$.

If $m$ is even, $c_m$ receives $c_{m-1}$ under $Y^*$. Choosing $X^*$ at $P^m$ would give her $c_{m+1}$, her first choice at $P^{m-1}$,
and hence a profitable deviation from $P^{m-1}$. Choosing deficient
type $X$ at $P^m$ would leave her with her initial position, although reporting as
at $P^{m-1}$ would secure $c_{m-1}$, her first choice at $P^m$.
Both choices violate SP. The rule must therefore choose $Y^*$.

If $m$ is odd, put $i=m$ and $j=m+1$. Since $m$ and $k$ are odd
and $m\le k-1$, we have $j\le k-1$. Thus $c_{j+1}$ is an incumbent,
and $c_k$ keeps her report with $v$ first at both intermediate
profiles below, so the vacancy remains assigned. We use two intermediate
profiles. Starting from $P^{m-1}$, let $c_j$ refuse $c_i$ by reporting
$c_{j+1}$ first, $c_j$ second, and $c_i$ below $c_j$.
Call the resulting profile $J$. At $P^{m-1}$, $Y^*$ gives
$c_j$ her second choice $c_i$. SP prevents her from receiving
$c_{j+1}$ at $J$, and IR excludes $c_i$, so she retains $c_j$.

At $J$, an unmatched incumbent can only have index $m-1$, $m$, or
$m+1$, omitting index zero when $m=1$. The same argument about retaining incumbents
as above excludes every other index. In particular, $c_{j+1}$
cannot be unmatched because $c_j$ still ranks $c_{j+1}$ first. The one
additional possibility,
$m+1$, arises because $c_{m+1}$ now refuses the claim from $c_m$.
Since the even-indexed $j=m+1$ is unmatched, parity requires exactly
one other unmatched incumbent. Type $X$ is impossible: its two
unmatched indices would have to be even followed by odd, whereas $j$
is the largest candidate and is even. Type $Y$ requires odd followed
by even, so the unmatched incumbents are exactly $c_i,c_j$.
The allocation is therefore $Y^*$ with their exchange omitted.

Next, at $J$, let $c_i$ switch to her report at $P^m$, putting
$c_{i-1}$ first and $c_j$ second, both above $c_i$. Call this profile
$K$. At $J$, both neighbors were acceptable to $c_i$ but she retained
her initial position, so SP prevents her from receiving either neighbor at
$K$. Thus she remains unmatched. At $K$ the same argument
restricts unmatched incumbents to $c_i,c_j$, so parity forces both to
be unmatched and again forces $Y^*$ with their exchange omitted.

Finally, restore $c_j$'s original report. This gives exactly $P^m$.
The rule cannot choose $X^*$ there, since it would give $c_j$
the object $c_{j+1}$, a profitable deviation from $K$. It cannot
choose deficient type $Y$ either: $c_i$ would then retain her
initial position at $P^m$, but reporting as at $P^{m-1}$ would secure the
acceptable position $c_j$. The remaining possibility is $Y^*$.
This completes the induction.

Hence the rule chooses $Y^*$ at $P^{k-1}$. Under the reversal
$c_t\mapsto c_{k+1-t}$, the profiles $P^k$ and $P^{k-1}$ play the
roles of $P^0$ and $P^1$, and $X^*$ plays the role of $Y^*$.
The odd-$m$ step with $m=1$ therefore shows that choosing $X^*$ at
$P^k$ forces $X^*$ at $P^{k-1}$, a contradiction. This proves
necessity for every $k$.
\end{proof}

\subsection{Computation and limits on exchange length}\label{app:policy}

Recall that $\Acore(P;\succ)$ denotes the core at preference profile $P$
under priorities $\succ$.

\begin{proposition}\label{prop:certifying}
Writing $n=|\mathcal{A}|+|\mathcal{V}|$ and $m=|E|$, the algorithm can be implemented in $O((n+m)^2)$ time and $O(n+m)$ space. On an arbitrary input, test the conclusion of \autoref{lem:safe} before every vacancy finalization. The resulting procedure either returns an allocation in $\Acore(P;\succ)$ or returns a conflict cycle.
\end{proposition}
\begin{proof}
We store preference lists, priority ranks, active flags, and rejection
flags. Each sweep computes pointers and rejections in $O(n+m)$ time.
If a rejection occurs, we charge the sweep to a newly rejected edge.
Otherwise, we check for departures and pointer cycles in Step~2. If
Step~2 does not apply, we follow each applicant's pointer path to the
vacancy at its end, saving computed endpoints for reuse. We then form
the dependency graph and find its sink components with a linear-time
strongly connected component algorithm. Each sweep without a rejection
removes at least one applicant. In total, there are at most $m$
rejection sweeps and $|\mathcal{A}|\le n$ departure sweeps, giving the
stated bounds.

Suppose a sink component fails the safety check: its vacancy $v$ ranks
an active applicant $a$ above its tentative recipient, but $a$ is not
held by a vacancy within the component. The link $v\to r(a)$ keeps $r(a)$ in the component,
so $a$ does not point to a vacancy. Since no newcomer points to
$\varnothing$ and no incumbent to her own position at this stage,
$a$ is an incumbent pointing to another incumbent's position $j$.

In the expanded dependency graph from \autoref{lem:safe}, there is a path
$v\to a\to j\to\cdots\to r(a)$. Since $r(a)$ can reach $v$ in the
dependency graph, expanding those links gives a return path to $v$.
Thus the arc $a\to j$ lies within one strongly connected component.
A shortest path from $j$ to $a$, preceded by $a\to j$, is a directed
cycle. No pointer cycle remains, so this cycle contains a vacancy. It
enters every vacancy from the applicant it holds and leaves toward a higher-priority
applicant, making it the required conflict cycle. Breadth-first search
extracts this cycle in linear time.

If every finalization passes, the proof of \autoref{lem:existence}
applies, since the absence of conflict cycles was used there only
to ensure safe finalization. The run therefore returns a core allocation
for the submitted profile.
\end{proof}

The polynomial-time bound in \autoref{prop:certifying} applies to a
run at one preference profile. A successful run need not certify that
the policy is conflict-free. For instance, if every applicant ranks her
outside option first, all leave in Step~2 under any policy. One can
test the policy itself by checking both orientations of every cycle
containing a vacancy and an exchange edge. We give no polynomial-time
bound for that test.

\begin{proof}[Proof of \autoref{prop:pairwise}]
If the policy has a conflict cycle, \autoref{lem:necessity} rules out
any SP, IR, and first-choice PR rule. We next show that on any simple
incumbent cycle, every such rule must execute the entire cycle at some
profile.

Orient a simple cycle $C=(c_0,\ldots,c_{m-1})$ of $m\ge3$ incumbents,
with indices read modulo $m$. At $P$, let each $c_t$ rank $c_{t+1}$
first, $c_{t-1}$ second, and her initial position third. The remaining
rankings are arbitrary. By IR each of the $m$ members receives one of
the $m$ positions on $C$, so injectivity assigns all these positions
within $C$. A member who moves receives one of her two neighbors'
positions, even if $C$ has chords. Hence any exchange involving three
or more members follows the entire cycle in one of its two directions.
All other trades on $C$ are swaps between neighbors.

The reverse exchange would deny each member her successor's position,
although the successor receives the claimant's position. This violates
first-choice exchange-pair rationality. A member cannot retain her
initial position either. Her predecessor ranks that position first,
and the member prefers her predecessor's position to her own. Thus
the rule must either give every member her successor's position or
pair all members with neighbors. If $m$ is odd, no pairing covers $C$.

Suppose $m$ is even and the rule pairs all members. Choose a member
$i$ paired with her predecessor. Let her keep her successor first and
report her own position second, ranking all other objects below it.
IR still assigns all positions on $C$ within $C$ and now excludes the
reverse exchange for $i$. If the rule executes the forward exchange,
$i$ receives her successor's position. Otherwise the members who move
trade in pairs. Every member other than $i$ must move, since her
predecessor still ranks her position first and she still prefers her
predecessor's position to her own. As $m$ is even and each swap
covers two members, $i$ must move too. By IR she can receive only her
successor's position when she moves. Thus the report gives $i$ her
true first choice instead of her predecessor's position, contrary to
SP. This proves that the forward exchange is forced at $P$, so a
cycle with $m>L$ rules out the proposed limit.

Conversely, suppose the policy is conflict-free and every simple
incumbent cycle has at most $L$ members. By \autoref{lem:structure},
each feasible exchange with three or more members follows a simple
incumbent cycle. Since $L\ge2$, every feasible allocation meets the
limit. By \autoref{thm:uniqueness}(i), PTR satisfies SP, IR, and PR.
\end{proof}

In the complete
housing market, \citet[Corollary~2]{EY2026} show that capping exchange
cycles below the size of the market rules out a strategy-proof rule
satisfying their first- and second-choice conditions.

Suppose incumbents are divided into categories and two distinct
incumbents may exchange if and only if they belong to the same category.\footnote{Category restrictions of this kind appear in public-sector
mutual-transfer rules. Turkey's Ministry of Family and Social Services
limits requests to otherwise eligible staff sharing a service region,
title, and staffing category (see its
\href{https://www.aile.gov.tr/media/301285/2026-yer-degisikligi-duyurusu.pdf}{2026 internal-transfer notice},
p.~4, quoting Art.~20(2) of the appointment and transfer regulation).
In Spain's central government, Art.~38 of the
\href{https://www.boe.es/buscar/act.php?id=BOE-A-2019-7414}{IV Convenio \'Unico}
requires exchanged posts to have the same professional group,
occupational family and/or specialty, and contractual regime.
Both systems also require administrative approval. The category
assumption captures the category restrictions while abstracting from
these additional approval requirements.} A category of $k\ge 3$ incumbents contains a cycle of length $k$, and every
exchange stays within a category. By \autoref{prop:pairwise}, a conflict-free
policy therefore admits an SP, IR, and first-choice PR rule limiting each
exchange to $L$ incumbents exactly when each category has at most $L$ members.

\subsection{Reassignment of released positions}\label{app:chains}

A \emph{reuse rule} relaxes withdrawal while retaining the other
requirements for an allocation. An incumbent who takes an announced
vacancy or a released position releases her own for reassignment in
the same round. Positions kept by their incumbents or exchanged in
cycles are never released. A released position is open to incumbents
compatible with its owner and a specified set of newcomers.

A newcomer's preferences may be any strict order over her outside option,
her eligible announced vacancies, and the positions she is eligible to
receive upon release. Each released position has a strict priority
order over its eligible applicants.
A rule \emph{respects priority at released positions} if, whenever an
eligible applicant prefers a released position to her assignment, it goes
to someone who outranks her there. A released position may remain
unassigned if no eligible applicant prefers it to her assignment.

The construction for two incumbents uses the following property of
sequential choice. At her turn, each newcomer chooses her favorite
remaining object or remains unmatched.

\begin{lemma}\label{lem:reuse-one-deletion}
Fix a pool of objects and a priority order of newcomers. After they
choose in order, a decision based only on their choices either confirms
the assignments or repeats the choices in the same order with one
predetermined object removed from the initial pool. Truthful reporting
is weakly dominant for each newcomer.
\end{lemma}
\begin{proof}
In the second run, each newcomer's menu is contained in her first menu
and lacks at most one object. This holds for the first newcomer. If a
newcomer makes the same choice in both runs, the difference between
the pools is unchanged. If her choices differ, her first choice was
the missing object. After she chooses again, the only object that can
be missing for the next newcomer is her new choice, if she took an
object. This proves the assertion by induction, including the option
to remain unmatched.

Fix the other reports. A newcomer's two menus depend only on earlier
choices in the respective runs, so neither depends on her report.
Confirmation gives her the best object in the larger menu. After a
rerun, the only strict improvement she could obtain is her truthful
first-run choice, if it is missing from her second menu. She can obtain
that object only in a confirmed first run. Choosing it in the first run
leaves all later first-run choices unchanged, so the decision still
calls for a rerun.
\end{proof}

\begin{proposition}[Reuse with priorities at released positions]\label{prop:reuse}\label{prop:released-priority}
Suppose there is at least one announced vacancy, all incumbents are
mutually compatible, and every applicant is eligible for every announced
vacancy. Every released position is open to every applicant other than
its initial holder. At every announced vacancy and released position,
eligible applicants are ranked according to a common priority order. An SP,
IR, and PR reuse rule that respects priority at
released positions exists if and only if there is at most one incumbent,
or there are exactly two incumbents and one ranks first among all
applicants.
\end{proposition}
Under withdrawal, the benchmark is implementable exactly when there is
at most one incumbent. Any two incumbents form a conflict cycle with a
vacancy, and with fewer than two incumbents there is no exchange edge
(\autoref{thm:main}).

\begin{proof}[Proof of \autoref{prop:reuse}]
For necessity, suppose there are two incumbents and a newcomer ranks
above both. Name the newcomer Carol and the incumbents Ann and Bob,
with Ann ahead of Bob. Choose one vacancy $v$ and use the preferences
in the example in \bodysecref{sec:chains}. Ann ranks Bob's position
first, $v$ second, and her own position third. Bob ranks $v$ first,
Ann's position second, and his own position third. Carol ranks only
Bob's position above her outside option. Each of the three ranks all
other objects below her outside option, and every other applicant ranks
her outside option first. By IR, only Ann or Bob can receive $v$, and Bob claims it if it
is left empty. If Ann receives $v$, she prefers Bob's position,
and Bob weakly prefers hers to his assignment, so their pair blocks.
If Bob receives $v$, Carol must receive his released position. Ann
then keeps her own position despite preferring $v$ and outranking
Bob there. No IR allocation satisfies both PR and priority at
released positions.

If there are at least three incumbents, select any three and name them
Carol, Ann, and Bob in decreasing priority order. Use the same profile,
with Carol's initial position as her outside option. No one else wants
her position. If Bob takes $v$, Carol must still receive his released
position, and Ann has justified envy at $v$. If Ann takes $v$, she and
Bob still block. Leaving $v$ empty violates Bob's claim. IR excludes
all other recipients of $v$. Thus the same impossibility applies.

With no incumbents, let newcomers choose in priority order. With one,
use applicant-proposing DA on all objects, ranking the incumbent first
at her own position and using the common order among all other claimants.
Stability guarantees the incumbent at least her own position because
she has first priority there. It also protects the common order at
every announced vacancy and, if she moves, at her released position. DA is SP
\citep{DF1981,Roth1982MOR}.

Now let Ann rank first overall and Bob be the other incumbent. Write
$a,b$ for their positions and $V$ for the announced vacancies. If Ann
ranks $a$ or a vacancy first, assign it to her. Apply the one-incumbent
rule to the remaining applicants, including $a$ among the available
objects precisely when Ann took a vacancy.

Suppose Ann ranks $b$ first. Let $U=V\cup\{a\}$. Tentatively let the
newcomers ahead of Bob choose in order from $U$. Offer Bob both
initial positions and the vacancies left after these choices, even if
a newcomer tentatively took $a$. His choice determines the outcome.
\begin{enumerate}[label=(\roman*),nosep]
\item If Bob takes a vacancy, give Ann $b$ and confirm the tentative
choices. Both positions are released. Ann has first priority at $b$.
\item If Bob takes $a$, execute the swap. Repeat the newcomers' choices
from $U\setminus\{a\}$ since the swap protects both positions.
\item If Bob keeps $b$, give Ann her favorite $d$ in $U$. Repeat the
newcomers' choices from $U\setminus\{d\}$. This includes $a$ exactly
when Ann releases it.
\end{enumerate}
In all three cases the newcomers below Bob choose in order from the
remaining available objects.

Ann receives her first choice except when Bob keeps $b$ after she
requests it. Then her best alternative is $d$ and no report can obtain
$b$ for her. For fixed reports by the others, Bob either participates
in the one-incumbent rule or obtains his favorite from a fixed menu.
A newcomer below Bob has a fixed menu when Ann requests $b$. For
newcomers ahead of Bob, the critical fact is that Bob always has access
to \emph{both} initial positions. If he takes no vacancy, his report
therefore determines which of (ii) and (iii) occurs, independently of
the newcomers' reports. Holding the incumbents' reports fixed, a rerun
deletes one predetermined object, either $a$ or $d$.
\autoref{lem:reuse-one-deletion} establishes SP for these newcomers.
When Ann initially chooses $a$ or a vacancy, SP for the remaining
applicants follows from DA.

Each applicant can keep her outside option. When Ann chooses $a$ or a
vacancy, DA protects priority among the remaining applicants. When she
requests $b$, newcomers choose among the remaining vacancies and
released positions in priority order in the final run. Ann obtains $b$
or, if Bob keeps it, her favorite in $V\cup\{a\}$. Bob is
offered both positions and all vacancies left by higher-ranked
newcomers in (i). In (ii) and (iii), the vacancies they leave in the
final run form a subset of those Bob was offered, by
\autoref{lem:reuse-one-deletion}. He could also have chosen $a$.
Thus the rule respects priority and fills every vacancy or released
position that an eligible applicant prefers to her assignment. For exchange-pair
rationality, Ann either prefers her assignment to $b$, receives $b$
after Bob has chosen with $a$ available, or is refused because Bob
strictly prefers keeping $b$ to taking $a$.
\end{proof}

\fi 
\end{document}